\newif\ifpdflatex    
\pdflatextrue           

\documentclass[fleqn,usenatbib,useAMS,twocolumn]{aastex}
\usepackage{amsmath,esint}
\usepackage{url,aasmacros}
\usepackage{natbib}
\usepackage{soul, CJK}
\usepackage{multirow}
\usepackage{tablefootnote}

\usepackage{bm}
\usepackage{xspace}
\usepackage{color}
\usepackage{enumitem}
\usepackage{booktabs}

\usepackage{graphicx}	
\usepackage{amsmath}	
\usepackage{amssymb}	
\usepackage{bm}		
\usepackage{url}
\usepackage{amsbsy}
\usepackage{xspace}
\usepackage{hyperref}
\usepackage{longtable}
\usepackage[colorinlistoftodos]{todonotes}
\usepackage[flushleft]{threeparttable}

\usepackage[T1]{fontenc}
\usepackage{ae,aecompl}

\def\lesssim{\mathrel{\hbox{\rlap{\hbox{\lower5pt\hbox{$\sim$}}}\hbox{$<$}}}}
\def\gtrsim{\mathrel{\hbox{\rlap{\hbox{\lower5pt\hbox{$\sim$}}}\hbox{$>$}}}}

\usepackage{upgreek}                                                  
\usepackage{xspace}                                                       
\newcommand{\um}{$\upmu$m\xspace}            
\def\apjl{ApJ}%
\def\apjs{ApJS}%
\def\aap{A\&A}%
\shorttitle{RCB stars from PGIR}
\shortauthors{Karambelkar et al.}

\begin{document}
\title{Census of R Coronae Borealis stars I: Infrared light curves from Palomar Gattini IR}
\author[0000-0003-2758-159X]{Viraj R. Karambelkar}
\email{viraj@astro.caltech.edu}
\affiliation{Cahill Center for Astrophysics, California Institute of Technology, Pasadena, CA 91125, USA}

\author{Mansi M. Kasliwal}
\affiliation{Cahill Center for Astrophysics, California Institute of Technology, Pasadena, CA 91125, USA}

\author{Patrick Tisserand}
\affiliation{Sorbonne Universites, UPMC Univ. Paris 6 et CNRS, UMR 7095, Institut d’Astrophysique de Paris, IAP, 75014 Paris, France}

\author{Kishalay De}
\affiliation{Cahill Center for Astrophysics, California Institute of Technology, Pasadena, CA 91125, USA}

\author[0000-0003-3768-7515]{Shreya Anand}
\affiliation{Cahill Center for Astrophysics, California Institute of Technology, Pasadena, CA 91125, USA}

\author{Michael C. B. Ashley}
\affiliation{School of Physics, University of New South Wales, Sydney, NSW 2052, Australia}

\author{Alex Delacroix}
\affiliation{Caltech Optical Observatories, California Institute of Technology, Pasadena, CA 91125, USA}

\author{Matthew Hankins}
\affiliation{Arkansas Tech University, 203 W O St, Russellville, AR 72801}

\author{Jacob E. Jencson}
\affil{University of Arizona, Steward Observatory, 933 N. Cherry Avenue, Tucson, AZ 85721, USA}

\author{Ryan M. Lau}
\affil{Institute of Space and Astronautical Science, Japan Aerospace Exploration Agency, 3-1-1 Yoshinodai, Chuo-ku, Sagamihara,\\ Kanagawa 252-5210, Japan}

\author{Dan McKenna}
\affil{Caltech Optical Observatories, California Institute of Technology, Pasadena, CA 91125, USA}

\author{Anna Moore}
\affil{Research School of Astronomy and Astrophysics, Australian National University, Canberra, ACT 2611, Australia}

\author{Eran O. Ofek}
\affil{Department of Particle Physics and Astrophysics, Weizmann Institute of Science, Rehovot 76100, Israel}

\author[0000-0001-7062-9726]{Roger M. Smith}
\affil{Caltech Optical Observatories, California Institute of Technology, Pasadena, CA 91125, USA}

\author[0000-0002-4622-796X]{Roberto Soria}
\affil{College of Astronomy and Space Sciences, University of the Chinese Academy of Sciences, Beijing 100049, China}
\affil{Sydney Institute for Astronomy, School of Physics A28, The University of Sydney, Sydney, NSW 2006, Australia}

\author{Jamie Soon}
\affil{Research School of Astronomy and Astrophysics, Australian National University, Canberra, ACT 2611, Australia}

\author[0000-0002-1481-4676]{Samaporn Tinyanont}
\affil{University of California Santa Cruz, 1156 High St, Santa Cruz, CA 95064}

\author{Tony Travouillon}
\affil{Research School of Astronomy and Astrophysics, Australian National University, Canberra, ACT 2611, Australia}

\author[0000-0001-6747-8509]{Yuhan Yao}
\affiliation{Cahill Center for Astrophysics, California Institute of Technology, Pasadena, CA 91125, USA}


\begin{abstract}
We are undertaking the first systematic infrared (IR) census of R Coronae Borealis (RCB) stars in the Milky Way, beginning with IR light curves from the Palomar Gattini IR (PGIR) survey. PGIR is a 30\,cm $J$-band telescope with a 25\,deg$^{2}$ camera that is surveying 18000\,deg$^{2}$ of the northern sky ($\delta>-28^{o}$) at a cadence of 2\,days. We present PGIR light curves for 922 RCB candidates selected from a mid-IR color-based catalog \citep{Tisserand2020}. Of these 922, 149 are promising RCB candidates as they show pulsations or declines similar to RCB stars. Majority of the candidates that are not RCB stars are either long period variables (LPVs) or RV-Tauri stars. We identify IR color-based criteria to better distinguish between RCB stars and LPVs. As part of a pilot spectroscopic run, we obtained NIR spectra for 26 out of the 149 promising candidates and spectroscopically confirm 11 new RCB stars. We detect strong He I $\lambda 10830$ features in spectra of all RCB stars, likely originating within high velocity (200-400 km-s$^{-1}$) winds in their atmospheres. 9 of these RCB stars show $^{12}$C$^{16}$O and $^{12}$C$^{18}$O molecular absorption features, suggesting that they are formed through a white dwarf merger. We detect quasi-periodic pulsations in the light curves of 5 RCB stars. The periods range between 30-125 days and likely originate from the strange-mode instability in these stars. Our pilot run results motivate a dedicated IR spectroscopic campaign to classify all RCB candidates.\\ 

\end{abstract}

\section{Introduction}
\label{sec:intro}
R Coronae Borealis (RCB) stars form a distinct class of variable stars. These stars are notable for their extreme photometric variations and characteristic chemical compositions. They are characterised by deep, rapid declines in their brightness ($\approx9$ mag in $V$ band, \citealt{Clayton1996}), that can last for hundreds of days before they emerge from the low state back to initial brightness. RCB stars also belong to the class of hydrogen-deficient stars, with helium being the most abundant element in their atmospheres, followed by carbon and nitrogen \citep{Asplund2000}. It is debated whether these objects originate in a white dwarf (WD) merger \citep{Webbink84} or from a final helium shell flash in an evolved low mass star \citep{Iben1996}. RCB stars are thus potential low mass counterparts of type Ia supernovae in the double-degenerate (DD) scenario \citep{Fryer2008}.



Photometric and spectroscopic properties of RCB stars have been studied extensively. While at maximum light, they are known to pulsate with periods between 40-100 days and amplitudes of few tenths of a magnitude \citep{Lawson1996}. The stars can then undergo mass loss episodes that eject ``puffs" of dust around them \citep{Feast1997a}. If the dust is ejected along the line of sight to a star, its brightness decreases rapidly. The dust is eventually blown away by radiation pressure and the star rises back to its initial brightness. Spectra of RCB stars taken at maximum light suggest that most of them are F-G type supergiants \citep{Iben1996} (although a few ``hot" RCB stars with T$>10000$K exist). RCB stars with effective temperatures $T_{\mathrm{eff}}<6800$K show spectra with absorption features of molecules such as CO, CN and C$_{2}$ \citep{Morgan2003,Tisserand2020}. Most RCB stars do not show any Hydrogen features in their spectra. However, a few exceptions exist. The stars V854Cen, VCrA, U Aquarii and DY Cen are relatively more hydrogen-rich and their spectra show H-Balmer lines and CH bands \citep{Kilkenny1989,Lawson1989}. During a photometric decline, the dust enshrouded RCB star shows a mostly featureless spectrum with few emission lines.

There are two contending theories to explain the formation of RCB stars - double degenerate (DD) and the final flash (FF). In the DD scenario, RCB stars are proposed to be remnants of the merger of a He-core WD and a CO-core WD. In the FF scenario, they are proposed to be the product of a final helium flash in the central star of a planetary nebula. The DD scenario is supported by the discovery of large abundances of $^{18}$O in atmospheres of some RCB stars \citep{Clayton2007,Garcia-Hernandez2009}. Using the CO bandhead in the near-infrared (NIR) spectra, these studies measured an $^{16}$O/$^{18}$O ratio of order unity in several RCB stars. For comparison, this ratio is $\approx$500 in the solar system \citep{Geiss2002}. Such high quantities of $^{18}$O can be produced during a white dwarf merger \citep{Clayton2007,Jeffery2011}. In contrast, there is no such model in the FF scenario that can explain the $^{18}$O overabundance \citep{Garcia-Hernandez2009}. However, the FF scenario is supported by the detection of Li and $^{13}$C in few RCB stars \citep{Asplund2000,Rao2008}. Thus, the DD scenario could account for most of the RCB stars while a small fraction, those without an $^{18}$O overbundance, may be formed through the FF channel. 
There is a big discrepancy between the number of known and predicted Galactic RCB stars. 117 Galactic RCB stars are currently known \citep{Tisserand2020}, while the total number is expected to be much larger. Assuming a DD origin with a WD merger rate of $\approx10^{-2}$ yr$^{-1}$, \cite{Clayton12} estimate that there are $\approx5400$ RCB stars in the Milky Way. This number is consistent with that extrapolated from the RCB population of the Large Magellanic Cloud \citep{Alcock2001}. From a campaign to identify Galactic RCB stars using mid-IR colors, \citet{Tisserand2020} estimate the total number to be $\approx380-550$. This suggests a lower WD merger rate of $\sim10^{-3}$ yr$^{-1}$, which is consistent with theoretical estimates made from population synthesis via the DD channel \citep{Ruiter2009,Karakas2015}. An accurate estimate of the number of RCB stars is important to shed further light on their progenitors. In the DD scenario, this number will provide an independent probe of the rate of mergers of white dwarf binaries. This will be particularly useful as close white-dwarf binaries will be important gravitational wave sources for LISA \citep{Amaro-Seoane2017,Burdge2020}.

Modern time domain surveys provide an attractive avenue for resolving this discrepancy. The high cadence, long baseline photometric observations are ideal to flag RCB stars (from their declines). These can be further followed up spectroscopically to study their chemical compositions. Such studies have been carried out in the past with data from optical surveys such as the Massive Compact Halo Object project (MACHO, \citealt{Alcock2001,Zaniewski2005}), EROS-2 \citep{Tisserand2004,Tisserand2008,Tisserand2009}, the Catalina Survey \citep{Lee2015}, All Sky Automated Survey for Supernovae (ASASSN, \citealt{Tisserand2013,Shields2019,Otero2014}) and the Zwicky Transient Facility (ZTF, \citealt{Lee2020}). In this paper, we use data from Palomar Gattini IR (PGIR), an infrared time domain facility to conduct the first near-infrared (NIR) search for RCB stars.


Palomar Gattini IR (PGIR, \citealt{De2020,Moore2019}) is a 25 sq. degree J-band camera on a 30 cm telescope located atop Mt. Palomar. PGIR was commissioned in September 2018 and commenced survey operations in July 2019. PGIR surveys the entire northern sky (18000 sq. degrees, $\delta > -28^{o}$) to a depth of $J\approx 16$ mag (AB) and a cadence of $\approx 2$ days. In the Galactic Plane, the limiting magnitude drops to $J\approx 14$ mag (AB) due to confusion noise. RCB stars are inherently brighter in the J band compared to the optical. 95$\%$ of all known Galactic RCB stars have 2MASS J band magnitudes brighter than 14 mag. Even during a photometric decline, the J band brightness decreases by $\approx3$ mag \citep{Feast1997a} as opposed to 9 mag in the optical. Of the known Galactic RCB stars, 66$\%$ are brighter than 11 mag in the J band. This puts a majority of the Galactic RCB stars above the sensitivity of PGIR. Additionally, a large number of RCB stars are expected to be located towards the Galactic Center, in regions of high dust-extinction \citep{Tisserand2020}. The extinction due to dust is significantly lower in the IR compared to optical wavelengths. This makes PGIR an ideal instrument to conduct a systematic search for Galactic RCB stars. In the first part of this search, we focus on PGIR $J-$ band light curves for objects in a pre-existing catalog of candidate Galactic RCB stars \citep{Tisserand2020}.

The remainder of this paper is structured as follows$-$ in Section \ref{sec:lc_pris}, we describe our source catalog and present PGIR $J-$band light curves of 922 objects from this catalog. We use the light curves to identify promising RCB candidates from contaminants such as long period variables (LPVs) and RV-Tauri stars. In Section \ref{sec:col_pris}, we use our lightcurve-based classifications to identify IR color-criteria to distinguish between RCB stars and LPVs to subsequently prioritise our spectroscopic followup. In Section \ref{sec:spec_follow} we present results from a pilot NIR spectroscopic campaign to identify new RCB stars from our list. In Section \ref{sec:progenitors} we analyse the NIR spectra and light curves of several RCB stars to derive their radial velocities, photospheric temperatures, $^{16}$O/$^{18}$O ratios, pulsation periods and discuss the implications of these quantities on their formation channels. We conclude with a summary of our results and future prospects in Section \ref{sec:summary}.

\section{IR lightcurve-based prioritization}
\label{sec:lc_pris}
Our source catalog comes from the \citet{Tisserand2020} list of 2194 Galactic RCB candidates selected based on their WISE colors. We performed forced aperture photometry using a 3 pixel ($\approx 13 \arcsec$) aperture on all J-band PGIR images since November 2018 at the locations of these candidates to generate light curves for each of them. The average cadence between observations is $\approx$ 2 days. The baseline of observations is $\approx$ 500 days, which corresponds to two observing seasons of the Galactic plane. Further details of the imaging and photometric pipeline can be found in \citet{De2020}. We note that the photometry for sources brighter than J$\approx8.5$ mag may not be accurate due to non-linearity effects in the detector.

Of the 2194 targets of interest, 1209 sources lie in the on-sky area covered by PGIR ($\delta \gtrapprox -28^{o}$). Of these, 287 are fainter than the detection limit of PGIR. The remaining 922 sources have J-band light curves. In addition to RCB stars, this list also contains several contaminants. These mainly include Miras and dust enshrouded RV-Tauri stars and TTauri stars (cf. Section 2.4 in \citealt{Tisserand2020}). Miras are long-period variables (LPVs) with periods longer than 150 days. RV-Tauri and TTauri stars show pulsations on timescales of a few weeks \citep{Grankin2007}. In this aspect, their light curves are similar to RCB stars at maximum light. However, unlike RCB stars, some RV Tauri stars show subsequent deep and shallow minima. Some also exhibit a long term periodic trend in addition to the shorter periodic variations \footnote{\url{http://ogle.astrouw.edu.pl/atlas/RV_Tau.html}}. To weed out these periodic impostors, we fit sinusoids to the J band light curves of all candidates using the GATSPY \citep{VanderPlas2015} implementation of the Lomb Scargle method \citep{Lomb76,Scargle82}. We visually examine the fits to determine the reliability of the derived period.

With this information in hand, we find that the 1209 candidates can be divided into the following priority groups based on their PGIR light curves $-$
\begin{itemize}[leftmargin=1pt]
    \item \textbf{Priority A} : The lightcurve shows sharp, deep declines or rises, or pulsations at timescales of few weeks, and resembles that of an RCB star. We identify 149 candidates in this category. We further divide this category into 3 sub-categories based on the trend shown by their light curves as of July 2020 $-$ 48 candidates that are \emph{declining}, 53 candidates that are \emph{rising} and 48 candidates that are \emph{pulsating} (with periods $\lessapprox100$ days). 
    \item \textbf{Priority B} : The lightcurve is ambiguous, not sufficient for any firm classification. There are 279 candidates in this category. Of these 279, 58 candidates show an overall rise in their brightness and 44 show an overall decline. 29 candidates have too large photometric errors to identify any trend. 148 sources show erratic photometric variations. 
    \item \textbf{Priority C} : The lightcurve shows no significant photometric evolution ($<\approx 0.1$ mag) for the last 500 days. There are 199 candidates in this category. 
    \item \textbf{Priority D} : There are no detections in PGIR data. 287 candidates belong to this category. 
    \item \textbf{Priority E} : The lightcurve is consistent with that of an RV Tauri or a TTauri star. 23 candidates belong to this category. 
    \item \textbf{Priority F} : The lightcurve is consistent with an LPV. There are 272 candidates in this category. 
\end{itemize}

    

\begin{figure*}
    \centering
    \includegraphics[width=\textwidth]{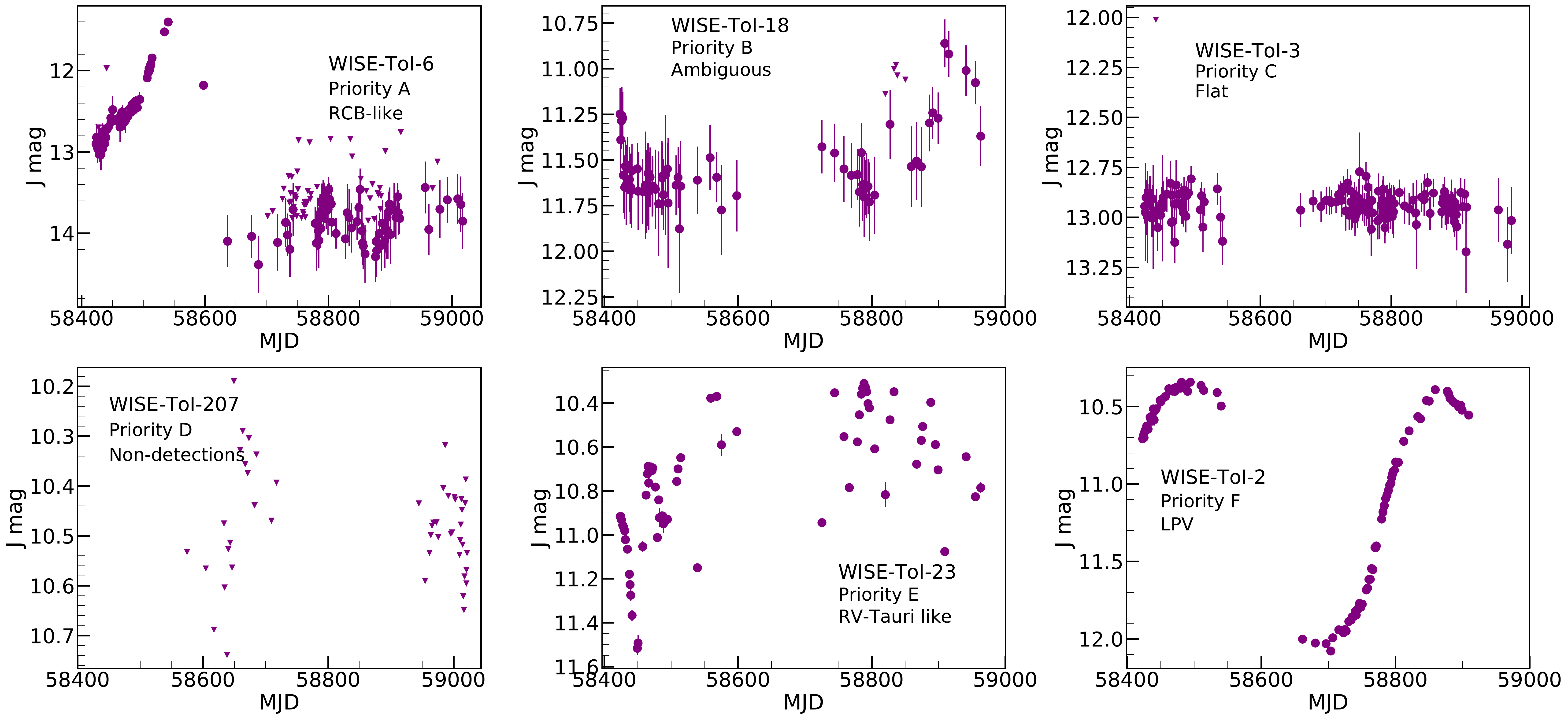}
    \caption{Examples of PGIR J-band lightcurves for candidates in each of the lc-based priority groups described in Section \ref{sec:lc_pris}. The solid dots represent detections and triangles represent 5-$\sigma$ upper limits. Of the 1209 candidates from \citet{Tisserand2020} catalog that have PGIR coverage, we assign 149 candidates Priority A (RCB-like), 279 candidates Priority B (ambiguous light curves), 199 candidates Priority C (flat light curves), 287 candidates Priority D (no detections), 23 candidates Priority E (RV Tauri-like) and 272 candidates Priority F (LPV-like). The Priority A group is expected to be rich in RCB stars, but the groups B, C  and D could also contain RCB stars. }
    \label{fig:lc_pris}
\end{figure*}

The groups are listed in order of priority for spectroscopic followup. Figure \ref{fig:lc_pris} show examples of lightcurves in each priority group. The Priority A group is expected to be rich in RCB stars. Some objects in this group could be LPVs that show non-sinusoidal variations. Priority B, C and D groups are also expected to contain RCB stars. Additional PGIR observations will help identify candidates with RCB-like declines or pulsations in these groups. Table \ref{tab:cands_pris} lists the lightcurve based priorities and sub-priorities for each of the 1209 candidates. 
\begingroup
\renewcommand{\tabcolsep}{2pt}
\begin{table*}
\begin{center}
\begin{minipage}{10cm}
\caption{Light curve (LC) based priorities, sub-classifications and IR color based priorities for spectroscopic followup of candidates in \citet{Tisserand2020} catalog. We assign color-based priorities to all 2194 candidates in the catalog, and lightcurve-based priorities to 1209 candidates that have PGIR coverage.}
\label{tab:cands_pris}
\begin{tabular}{cccccc}
\hline
\hline
{WISE} & {R.A} & {Dec} & {LC-based} & {LC-based} & {IR-color based}\\ 
{ToI} & {(deg)} & {(deg)} & {Priority} & {sub-class} & {priority}\\
\hline
1 & 2.1755 & 63.0093 & A & pulsating & 1-b\\
2 & 4.7942 & 52.0343 & F & & 1-a\\
3 & 7.3382 & 64.8141 &  C & & 1-a \\
4 & 9.7554 & 59.4681 &  B & erratic & 1-a\\
5 & 11.6178 & 58.9057 & C & & 1-a\\
6 & 12.0930 & 74.2992 & A  & declining & 1-b\\
... & & & & \\
\hline
\hline
\end{tabular}
\begin{tablenotes}
 \item This table is published online in its entirety in the machine-readable format.
\end{tablenotes}
\end{minipage}
\end{center}
\end{table*}
\endgroup

\section{IR color-based prioritization}
\label{sec:col_pris}

\begin{figure*}
    \centering
    \includegraphics[width=\textwidth]{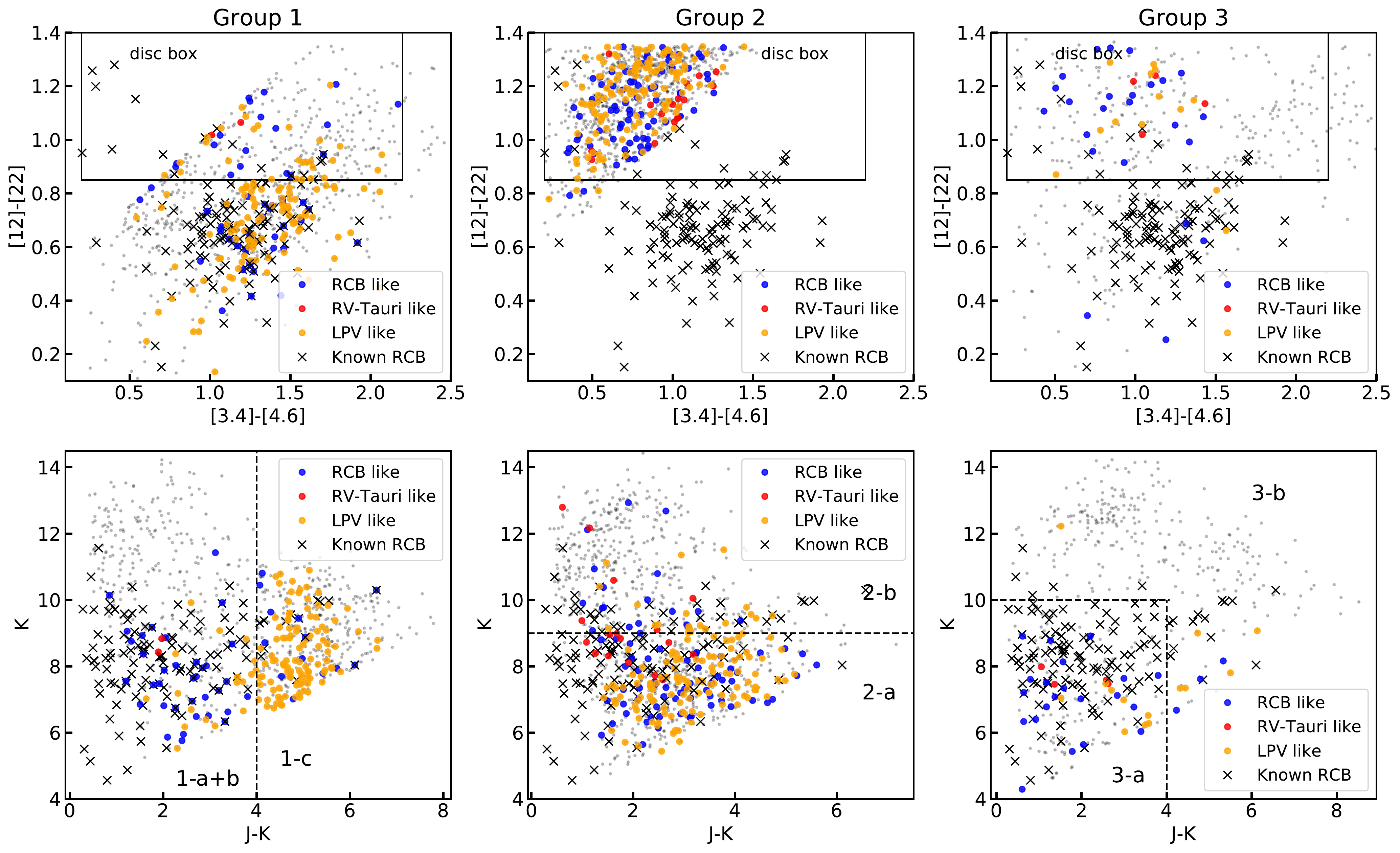}
    \caption{The WISE [3.4]$-$[4.6] vs [12]$-$[22] and 2MASS $K$ vs $J-K$ color-color diagrams for all candidates in the \citet{Tisserand2020} catalog. The left column shows the plots for candidates in Group 1, middle column shows candidates in Group 2 and the right column shows candidates in Groups 3. All candidates in each group are plotted as gray background points. On each color-color diagram, we highlight candidates that have light curves similar to RCB stars (Priority A, blue circles), RV Tauri stars (Priority E, red circles) and LPVs (Priority F, orange circles) present in the respective group. In addition, we also plot all known RCB stars as black crosses. We assign a new color based priority to each candidate based on its position in the K vs J$-$K diagram, as indicated. In the WISE color-color plots, we also indicate the ``disc-box" where RV-Tauri stars surrounded by dust discs are expected to lie \citep{Gezer2015}. In Group 1, J-K$= 4$ line separates LPVs from the known RCB stars and lc-Priority A candidates. We assign all candidates in Group 1 that have $J-K<4$ mag and lie outside the RV-Tauri disc box to ``Group-1-a" and those inside the disc box to ``Group-1-b". For Groups 2 and 3, we cannot identify a similar distinction between LPVs and RCB candidates. Instead, we use the positions of known RCB stars and our light curve based priority A candidates to prioritise these groups for spectroscopic followup (see text). We prioritise candidates in Group-1-a, Group-2-a and Group-3-a for spectroscopic followup.}
    \label{fig:phot_priority}
\end{figure*}

\begin{figure*}[htp]
    \centering
    \includegraphics[width=\textwidth]{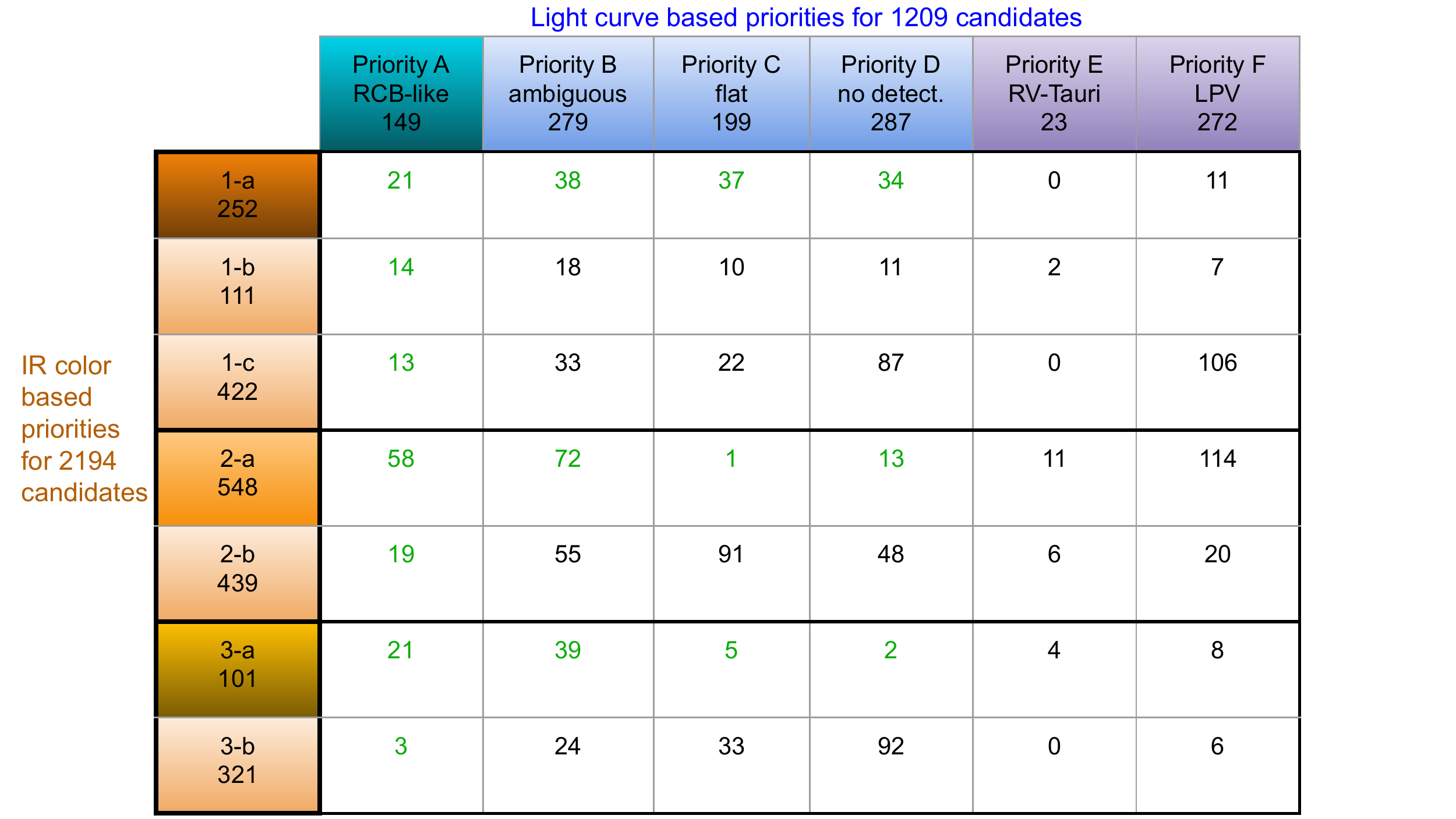}
    \caption{Schematic of light curve-based priorities and IR color based priorities for RCB candidates. We assign light curve based priorities (A$-$F) to 1209 of the 2194  candidates in \citet{Tisserand2020} catalog. We also divide the full catalog into 7 groups based on the IR colors of the candidates. In this figure, we indicate the overlap between each lightcurve based and IR color based group. For our campaign, we aim to observe all 149 candidates with lightcurve-priority A. We will also observe 241 additional candidates that have lc-Priorities B, C, D \emph{and} belong to IR-color Groups 1-a, 2-a and 3-a. We highlight these 390 candidates in green.}
    \label{fig:pri_schema}
\end{figure*}

In Section \ref{sec:lc_pris}, we prioritized 1209 of the 2194 candidates from \citet{Tisserand2020}'s catalog based on their PGIR light curves. Importantly, we identified 272 LPVs and 23 RV-Tauri stars from their light curves. Here, we explore the positions of these contaminants in the WISE and 2MASS color diagrams. We use this to identify color criteria that distinguish between RCB stars, LPVs and RV Tauri stars. We then assign a color-based priority to each of the 2194 candidates to optimise further NIR spectroscopic follow-up. This priority will be particularly useful for the 985 candidates that do not have PGIR coverage.


The WISE and 2MASS color-color diagrams were used previously by \citet{Tisserand2020} to divide their catalog into different groups (their Figure 4). Group 1 occupies the bottom-right region in the WISE [3.4]$-$[4.6] vs [12]$-$[22] diagram. This group is expected to contain the maximum number of RCB stars. Group 2 occupies the top-left region in the diagram and is expected to contain rare RCB stars that are surrounded by thin or cold dust shells (T$<500$ K). Finally, Group 3 comprises of objects that are reported with upper limits in at least one WISE band and objects that have prior classifications. We analyse the candidates in each of these groups separately \footnote{\citet{Tisserand2020} further divide Group 1 and 3 into two subgroups each}.


Figure \ref{fig:phot_priority} shows the WISE [3.4]$-$[4.6] vs [12]$-$[22] and 2MASS $K$ vs $J-K$ color-color diagrams for all 2194 candidates in the source catalog. We plot these diagrams separately for Groups 1, 2 and 3. All candidates in a group are plotted as gray background points. On each plot we highlight the candidates with lightcurve-based priority A (RCB-like, blue dots), priority E (RV-Tauri, red dots) and priority F (LPVs, orange dots). We also indicate the position of known RCB stars in these diagrams (black crosses).  In the [3.4]$-$[4.6] vs [12]$-$[22] diagram, we also mark the ``disk-box" region, which is expected to contain RV Tauri stars with dust disks around them \citep{Gezer2015}. We note that our lightcurve-Priority E (RV-Tauri) candidates lie in this region.

Based on these diagrams, we assign a color-based classification to each of the 2194 candidates as follows - 
\begin{itemize}[leftmargin=*]
    \item \textbf{Group 1} : This group has 785 candidates, and are expected to contain the maximum number of RCB stars. The left column in Fig. \ref{fig:phot_priority} shows the color-color diagrams for candidates in this group. These candidates lie in the bottom-right part of the [3.4]$-$[4.6] vs [12]$-$[22] diagram. We first note that the K vs J$-$K diagram can be used to separate most LPVs from RCB stars in this group. If we apply a cut of J$-$K $< 4$ mag, we eliminate 90$\%$ LPVs, while retaining 90$\%$ of known RCB stars and 73$\%$ priority A candidates. We assign all 422 candidates with J$-$K $>4$ mag the priority ``Group-1-c". For the few RCBs stars lying within this group, the 2MASS observations were likely obtained while they were in a deep decline phase, being, consequently, highly reddened. Of the remaining 363 candidates, 111 candidates lie in the RV Tauri disk box. We assign a priority ``Group-1-b" to these stars. We give the highest priority for spectroscopic follow-up to the remaining 252 candidates -- ``Group-1-a".
    
    
    \item \textbf{Group 2} : This group alone has 987 candidates, and is expected to have the highest contamination. The middle column in Fig. \ref{fig:phot_priority} shows the color-color diagrams for candidates in Group 2. These candidates lie in the top left part of the [3.4]$-$[4.6] vs [12]$-$[22] diagram. Although this group contains the largest number of candidates, very few known RCB stars (7 out of 117) lie in this region of the diagram. The disk box is not a particularly useful metric for prioritization as almost all of these candidates lie in the disk box. Unlike Group 1, no clear distinction can be made between the LPVs, known RCB stars and photometric RCB candidates in the $K$ vs $J-K$ diagram. If we apply a cut of $K<9$ mag, we retain 70$\%$ of known RCB stars and 75$\%$ of lc-priority A candidates. In absence of any other criteria, we prioritise the sources with $K<9$ mag over the remaining candidates. We assign 548 candidates with K$<$ 9 mag to the higher priority ``Group-2-a" and the remaining 439 candidates to the lower priority ``Group-2-b".
    
    \item \textbf{Group 3} : This group has 422 candidates. The last column in Fig. \ref{fig:phot_priority} shows the color-color diagrams for candidates in Groups 3. Based on the K vs J-K diagram, we prioritise sources that lie inside the box defined by $4<K<10$ and $J-K<4$ mag. This box contains 85$\%$ of known RCB stars, 88$\%$ of priority A candidates and only $50\%$ of LPVs. We assign 150 candidates that lie in this box priority ``Group-3-a" and the remaining 272 candidates ``Group-3-b".
\end{itemize}
We list these candidates with their color-based priorities in Table \ref{tab:cands_pris}. For spectroscopic followup, we prioritise candidates in Group-1-a, Group-2-a and Group-3-a in decreasing order. Group-1-b is also expected to contain RCB stars but is likely contaminated by RV-Tauri stars. The IR colors thus provide an additional metric to prioritise the candidates for spectroscopic followup. 

Our spectroscopic campaign focuses on candidates in the northern hemisphere that lie in the area covered by PGIR. First, we aim to spectroscopically follow-up all 149 candidates that have light curve-based Priority A. From the 765 candidates that have light curve-Priority B, C or D, we will follow-up only 241 candidates that belong to our top three IR-color based groups (Groups-1-a, 2-a and 3-a). This leaves us with a sample of 390 candidates for spectroscopic followup. The distribution of the candidates among the different priority groups is illustrated in Figure \ref{fig:pri_schema}.

\section{Pilot NIR Spectroscopic Followup}
\label{sec:spec_follow}
In Sections \ref{sec:lc_pris} and \ref{sec:col_pris}, we reprioritised candidates in the \citet{Tisserand2020} catalog based on their IR light curves and colors. We aim to spectroscopically followup the candidates in the top priorities to confirm their true nature. Traditionally, large scale spectroscopic campaigns for identifying RCB stars have made use of optical spectra. As noted in Section \ref{sec:intro}, RCB stars are significantly brighter in the NIR compared to the optical. Thus, NIR wavebands are better suited to efficiently observe large number of candidate RCB stars. In addition, NIR spectral features such as the $^{12}$C$^{16}$O and $^{12}$C$^{18}$O bandhead can provide useful diagnostics about the progenitors of these stars. Thus, a NIR spectroscopic campaign promises to be highly productive in identifying and characterising RCB stars. No such campaign has been conducted in the past. As a pilot for such a campaign, we obtained NIR spectra for eight known RCB stars and 31 of our top lightcurve-based priority candidates. Of the 31 candidates, 26 have lightcurve-priority A and five have lightcurve-priority B. Here, we present results from this pilot run. 

We used Triplespec \citep{Herter2008}, a medium resolution (R$\approx2700$) \emph{JHK} band spectrograph on the 200-inch Hale telescope at Mt. Palomar. We observed eight known RCB stars and 21 candidates on 2019 October 23, and three additional candidates on 2020 February 3. We further observed seven more candidates on 2020 June 28 with the Spex spectrograph (R$\approx1500$) on the NASA Infrared Telescope Facility (IRTF,\citealt{Rayner2003}) at Mauna Kea. The IRTF observations were acquired as part of program 2020A111 (PI: K. De). All spectra were extracted using the IDL package \texttt{spextool} \citep{Cushing2004}. The extracted spectra were flux calibrated and corrected for telluric absorption with standard star observations using \texttt{xtellcor} \citep{Vacca2003}.

\subsection{NIR spectral features of known RCB stars}
\label{sec:spfeats_known}
We obtained NIR spectra of eight known RCB stars (as listed on SIMBAD\footnote{\url{https://simbad.u-strasbg.fr}}) $-$ A0 Her, V391 Sct, NSV 11154, WISE J184158.40-054819.2 (WISE J18+), WISE J175749.76-075314.9 (WISE J17+), ASAS-RCB-21, WISE J194218.38-203247.5 (WISE J19+) and ASAS-RCB-20.

\begin{figure*}[htp]
    \centering
    \includegraphics[width=0.81\textwidth]{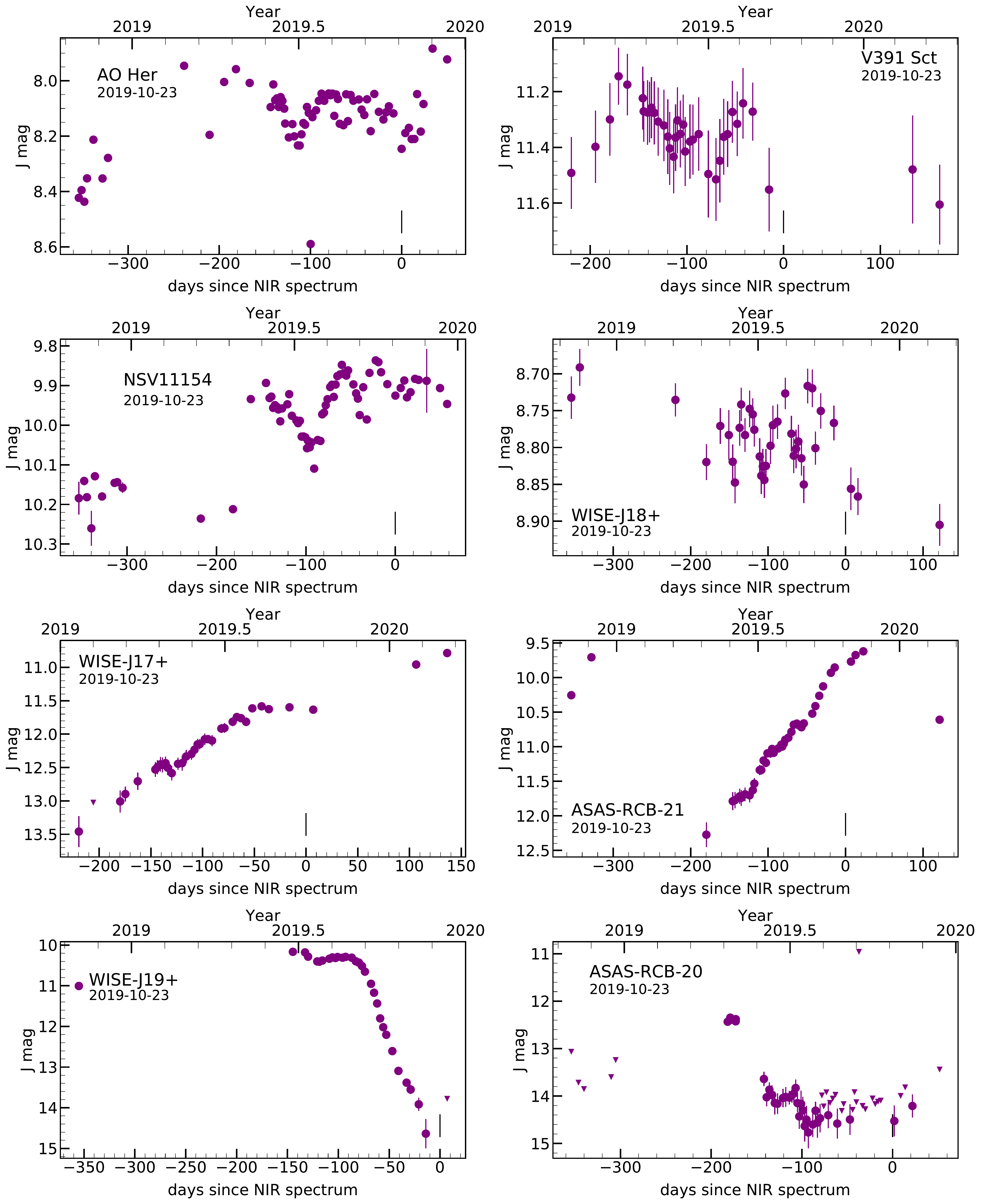}
    \caption{PGIR \emph{J} Band light curves for 8 known RCB stars. The black vertical lines at 0 days mark the epoch at which the NIR spectra were acquired (indicated in each plot). We note that photometry brighter than 8.5 mag may not be reliable due to non-linearity effects in the detector. The J-band magnitudes are in Vega system.}
    \label{fig:known_rcbs_lcs}
\end{figure*}

\begin{figure*}
    \centering
    \includegraphics[width=\textwidth]{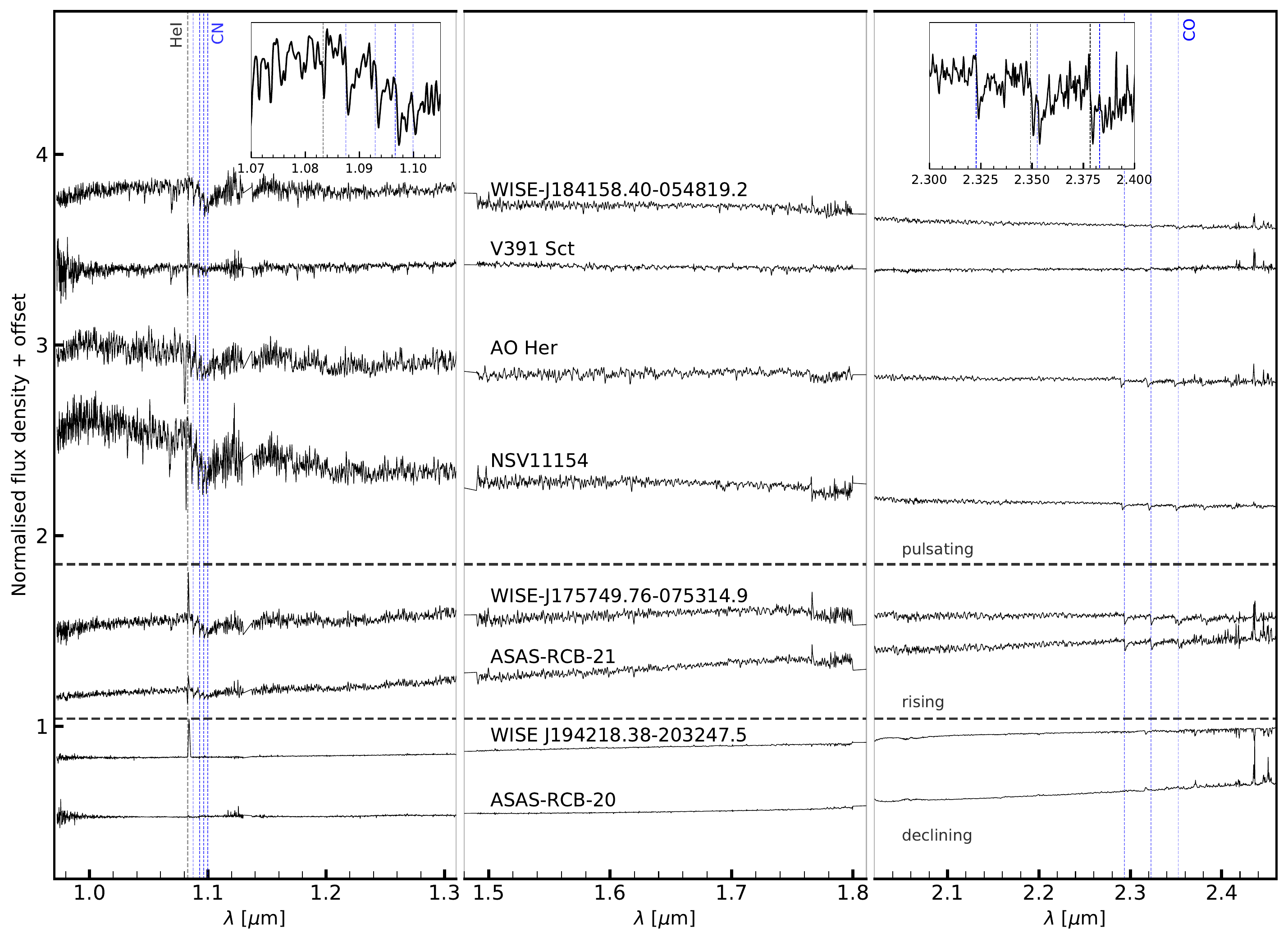}
    \caption{NIR spectra of 8 known RCB stars on 2019 November 23. The stars are grouped by their photometric phase - 4 pulsating, 2 rising and 2 fading. We also show a zoom in of the J-band and K-band spectra of the RCB star WISE-J18+. In the J-band inset plot, we mark the positions of the CN absorption bandhead in blue and He I absorption in gray. In the K-band inset, we mark the position of $^{12}$C$^{16}$O and $^{12}$C$^{18}$O absorption bandheads in blue and black respectively.}
    \label{fig:known_rcbs_spec}
\end{figure*}

Figure \ref{fig:known_rcbs_lcs} shows the \emph{J} band light curves of these stars. Seven of the known RCB stars have been reported as cold ($T_{\mathrm{eff}}<6800$K) $-$ AO Her (Tisserand P., private communication), NSV11154 \citep{Kijbunchoo2011}, ASAS-RCB-20, ASAS-RCB-21 \citep{Tisserand2013}, WISE-J17+, WISE-J18+ and WISE-J19+ \citep{Tisserand2020}. The star V391 Sct has been reported as a warm RCB by \citet{Tisserand2013}. Four out of the eight RCB stars $-$ V391Sct, NSV11154, AOHer and WISE-J18+ were at the maximum light phase when the spectra were obtained. All four stars exhibit quasi-periodic pulsations with periods between 40$-$140 days. The RCB stars WISE-J19+ and ASAS-RCB-20 were in the photometric minimum phase. WISE-J17+ and ASAS-RCB-21 were rising out of a photometric minimum towards maximum light. Figure \ref{fig:known_rcbs_spec} shows the NIR spectra of these stars, grouped by their photometric phase. We note the following characteristic features in these spectra $-$ 
\begin{itemize}[leftmargin=*]
    \item All RCB stars show He I ($\lambda 10830$) emission or absorption features. The two stars inside a photometric minimum show helium emission. The emission is strong for WISE-J19+ (just entering the minimum) and very weak for ASAS-RCB-20 (deep inside the minimum, featureless spectrum). The pulsating and rising stars show either a fully developed P-cygni profile or strong blueshifted absorption. We discuss the He I line profiles in more detail in Section \ref{sec:he_profiles}.
    \item The RCB stars inside a minimum show a mostly featureless reddened spectrum with He I emission. If the star has entered the minimum recently, other emission lines such as Si I and C$_{2}$ are detected. These features are absent if the star is deep inside the minimum. 
    \item The rising and pulsating RCB stars show a variety of atomic absorption lines. These include C I (most prominently at 1.0686 and 1.0688\um) along with Fe I, Si I and K I. 
    \item Of the rising and pulsating RCB stars, all but V391 Sct show the CN (1.0875, 1.0929, 1.0966 and 1.0999\um) and $^{12}$C$^{16}$O (2.2935, 2.3227, 2.3525, 2.3829, 2.4141\um) absorption bandheads. The presence of these molecular bandheads suggests that these stars have cold photospheres ($T_{\mathrm{eff}}<6800$ K). Importantly, we also detect $^{12}$C$^{18}$O absorption features in the K band of all these stars. We discuss the CO bandhead in more detail in Section \ref{sec:co_ratios}.
    \item None of the 8 RCB stars show any Hydrogen features in their NIR spectra.
\end{itemize}

\subsection{New RCB stars}
\label{sec:new_rcbs}

\begin{figure*}[t]
    \centering
    \includegraphics[width=\textwidth]{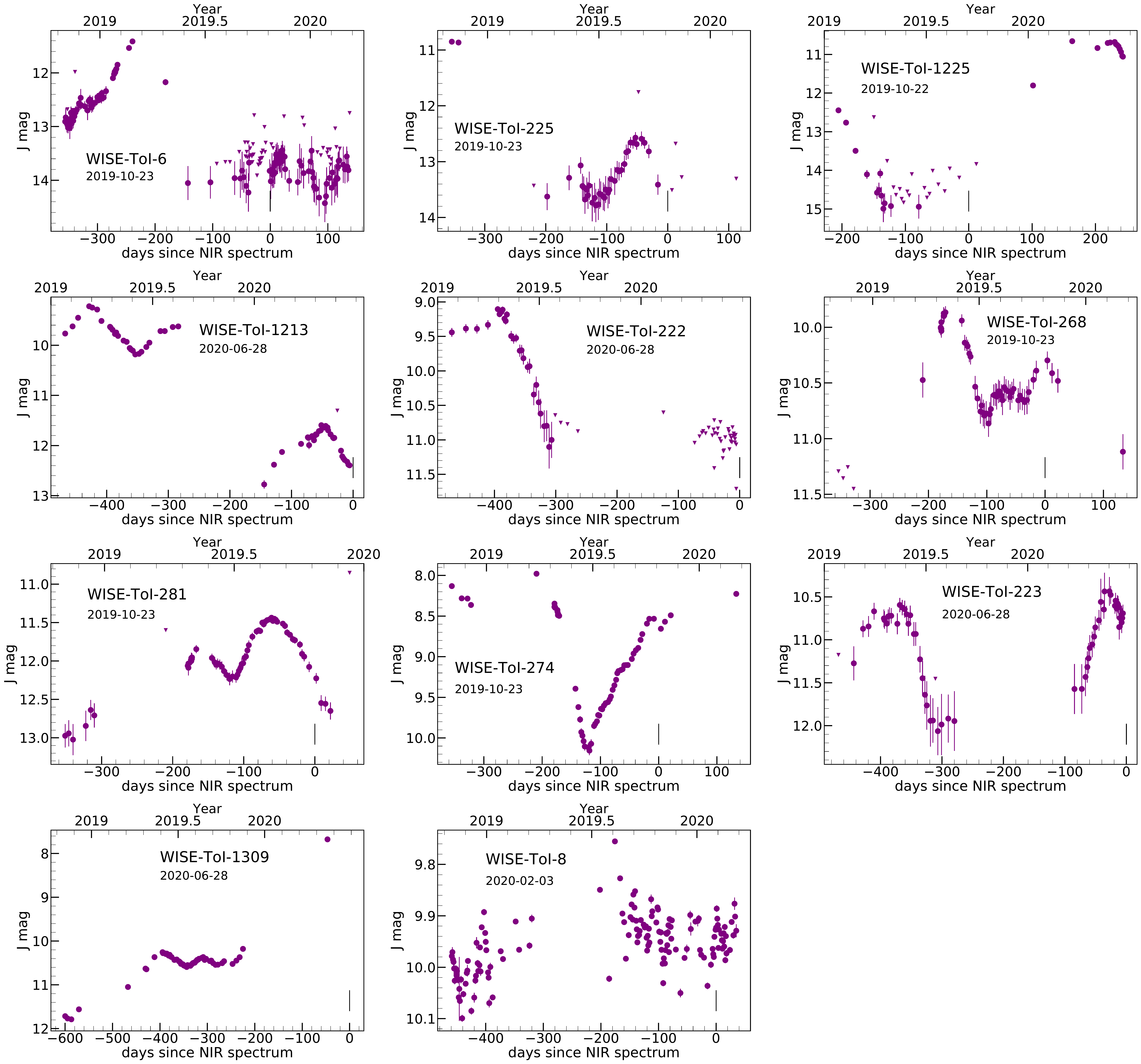}
    \caption{PGIR J band light curves of new RCB stars identified in this paper. The black vertical lines at 0 days mark the epochs at which the NIR spectra were acquired. For each RCB star, we also indicate the date when the spectrum was obtained. We note that photometry brighter than 8.5 mag may not be reliable due to non-linearity effects in the detector. All J-band magnitudes are in Vega system.}
    \label{fig:new_rcbs_lc}
\end{figure*}

\begin{figure*}[t]
    \centering
    \includegraphics[width=\textwidth]{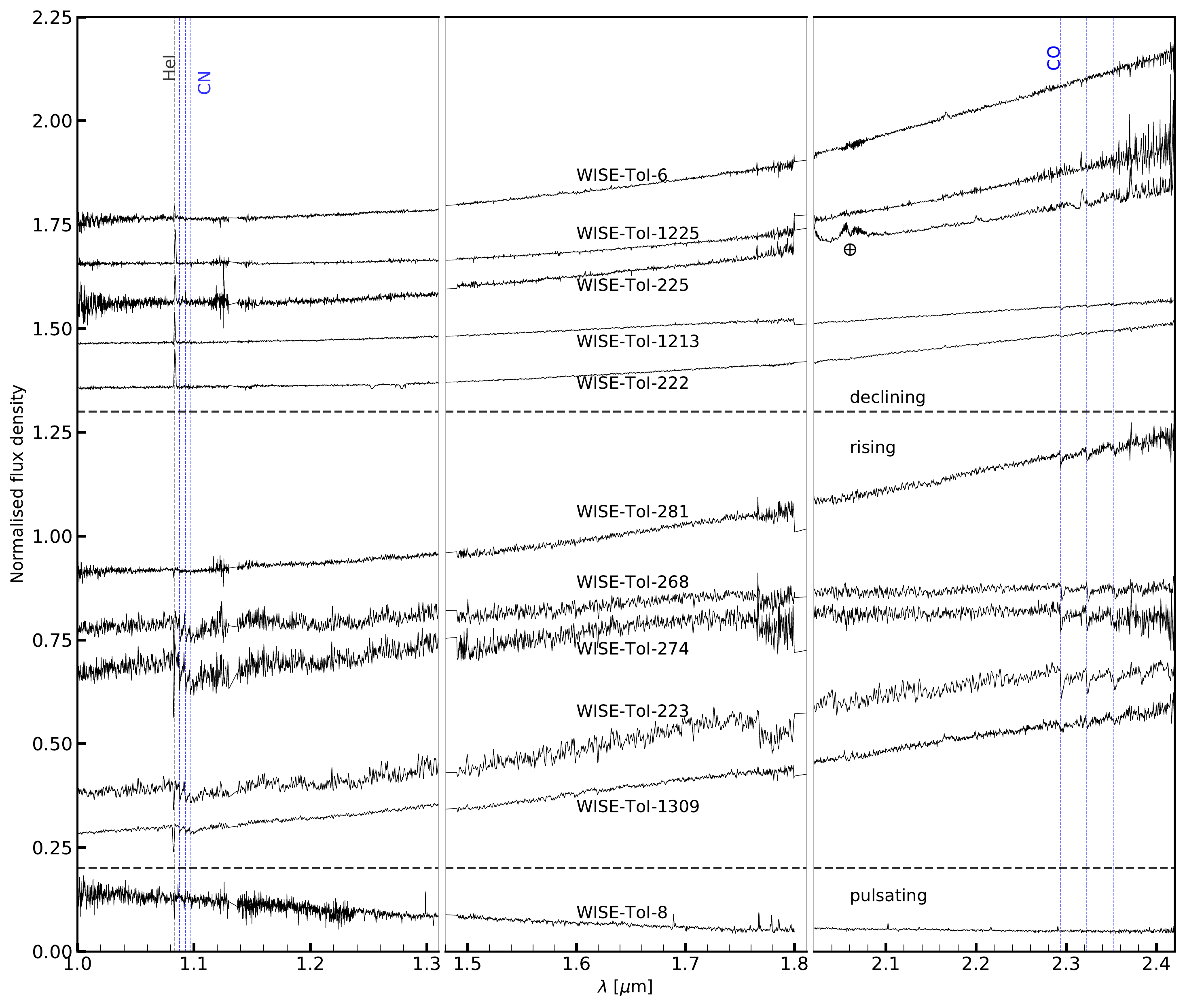}
    \caption{NIR spectra of new RCB stars identified in this paper. All RCB stars show He I $\lambda 10830$ features. 5 RCB stars that were in a deep decline show strong emission, while others show either fully developed p-cygni profiles or strong blueshifted absorption. 4 RCB stars show the CO absorption bandhead, suggesting that they are cold.}
    \label{fig:new_rcbs_spec}
\end{figure*}

\begingroup
\renewcommand{\tabcolsep}{2pt}
\begin{table*}
\begin{center}
\begin{minipage}{16cm}
\caption{Properties of newly identified RCB stars}
\label{tab:new_rcbs}
\begin{tabular}{lccccccc}
\hline
\hline
{WISE} & {WISE} & {RA} & {Dec} & {Light-curve} & {Temperature} &  A$_{\mathrm{V}}^{b}$ & Comments \\ 
{Name} & {ToI} & {(deg)} & {(deg)} &  {Phase$^{a}$} & {} & mag & \\
\hline
J004822.34+741757.4 & 6 & 12.0931 & 74.2993 &  Declining  & cold & 4.3 & \\
J005128.08+645651.7 & 8 & 12.8670 & 64.9477 &  Pulsating & hot & 1.2 & pulsating with period $\approx 30$ days\\
J181836.38-181732.8 & 222 & 274.6516 & -18.2925 & Declining & & 8.7 &  \\
J182010.96-193453.4 & 223 & 275.0457 & -19.5815 & Rising & cold & 4.9 & \\
J182235.25-033213.2 & 225 & 275.6469 & -3.5370 & Declining & & 6.5 & \\
J190813.12+042154.1 & 268 & 287.0547 & 4.3650 & Declining & cold & 6.8 & \\
J191243.06+055313.1 & 274 & 288.1795 & 5.8870 & Rising & cold & 7.7 & sharp rise out of a decline\\
J192348.98+161433.7 & 281 & 290.9541 & 16.2427 & Declining & cold & 22 & \\
J170552.81-163416.5 & 1213 & 256.4701 & -16.5713 & Declining &  & 1.3 & \\
J173737.07-072828.1 & 1225 & 264.4045 & -7.4745 & Declining &  & 2.8 &   \\
J185726.40+134909.4 & 1309 & 284.3600 & 13.8193 & Rising & cold & 2.4 & rising from 600 day long minimum \\
\hline
\hline
\end{tabular}
\begin{tablenotes} 
\item a : The photometric phase when the spectrum was obtained. 
\item b : Line of sight extinction, values taken from \citet{Schlafly11}
\end{tablenotes}
\end{minipage}
\end{center}
\end{table*}
\endgroup

We obtained NIR spectra for 31 RCB candidates $-$ 26 with lightcurve-priority A and five with lightcurve-priority B. None of the five priority B candidates show spectra resembling RCB stars. Of the 26 priority A candidates, we identify 11 new RCB stars using the spectral features described in Section \ref{sec:spfeats_known}. These stars are $-$ WISE-ToI-6, 8, 222, 223, 225, 268, 274, 281, 1213, 1225 and 1309. Table \ref{tab:new_rcbs} summarises the properties of these stars. Figure \ref{fig:new_rcbs_lc} shows the light curves of these stars and Figure \ref{fig:new_rcbs_spec} shows their spectra. Here, we discuss the properties of these newly identified RCB stars. We group the stars by their photometric phase at the time the spectra were obtained.

\subsubsection{Declining}
Seven of the 11 new RCB stars $-$ WISE-ToI-6, 222, 225, 268, 281, 1213 and 1225 were in a photometric minimum or were undergoing a photometric decline when the spectra were obtained. The light curves of WISE-ToI-6, 222, 225 and 1225 indicate that these stars were in a photometric minimum for more than a hundred days before the spectra were taken. These stars showed photometric declines of $\approx3$ mag in the J band and present featureless spectra with strong He I emission. WISE-ToI-268 and 281 show shallower declines ($\approx 1.5$ mag) and also show brief recoveries from the declines. Their spectra show strong blueshifted He I absorption and $^{12}$C$^{16}$O and $^{12}$C$^{18}$O absorption bands, suggesting that they are cold RCB stars. The lightcurve of WISE-ToI-1213 shows a complex photometric evolution where the star underwent shallow declines and recoveries of 1 mag \emph{and} a deep decline of $> 3$ mag. The spectrum obtained during this decline shows He I with a P-cygni profile. We also detect the $^{12}$C$^{16}$O absorption bandhead with traces of $^{12}$C$^{18}$O, suggesting that this is a cold RCB star.

We note that WISE-ToI-225, 222 and 281 lie in highly extincted regions of the Galaxy. The integrated line of sight extinction in their direction ($A_{V}$) is 6.5, 8.7 and 22 mag respectively \citep{Schlafly11}. Their 2MASS $J$ band magnitudes are 10.8, 9.0, 10.9 mag and PanSTARRS $g$ band magnitudes are 19.9, 20.2 and 22.4 mag respectively. These stars thus have $J-g \gtrapprox 10$ mag and illustrate the advantage that NIR searches offer over even very deep optical searches for RCB stars.

\subsubsection{Rising}
Three stars WISE-ToI-223, 274 and 1309 were rising out of a photometric decline when the spectra were obtained. ToI 223 was at constant minimum brightness for 200 days before it rose out of the minimum, brightening by 2.5 mag in J band. ToI 274 previously underwent a decline of 2 mag but recovered sharply without flattening at the minimum brightness. This variation is similar to the one observed with DYPer type stars \citep{Tisserand2009}. The J band lightcurve of ToI 1309 suggests that this star was in a minimum for at least 600 days before our spectrum was obtained. 
The spectra of these stars show strong blueshifted He I absorption, CN, $^{12}$C$^{16}$O and $^{12}$C$^{18}$O absorption features. ToI 274 also shows a small redshifted He I emission component.

ToI 223 and 274 have been listed as strong RCB candidates by \citet{Tisserand2020}. They note that these stars are the fifth and sixth brightest RCB stars in the 12\um band. They have WISE [12] = 1.0 and 1.2 mag respectively, indicative of a very bright circumstellar dust shell. The integrated Galactic extinction in their direction ($A_{V}$) is 4.9 and 7.7 mag respectively, suggesting that they lie in regions of high dust extinction. 

\subsubsection{Pulsating}
The lightcurve of WISE-ToI-8 was showing quasi-periodic variations when the spectrum was obtained. We identify a tentative period of $\approx 30$ days for this star. This star was photometrically identified as an RCB candidate on AAVSO and by the ASAS-SN survey \citep{Jayasinghe2018}, as it underwent photometric declines in 1999 and 2012. This star has not shown any obvious photometric declines at least for the last 500 days. The spectrum shows blueshifted helium absorption and no Hydrogen absorption lines. No CO or CN features are present, suggesting that this is not a cold RCB star. The spectrum shows several emission lines, resembling spectra of hot (T$_{\mathrm{eff}}>10000$ K) RCB stars. We identify CI emission at 1.68\um, CII emission at 1.784\um, FeII emission at 1.766 and 1.778\um and possibly Ne I emission at 1.7961\um. Similar to other hot RCB stars, the emission lines are double-peaked, which could indicate the presence of an equatorial outflow around the star \citep{DeMarco2002}. 

Our pilot campaign has demonstrated the utility of NIR spectra in searches for RCB stars. We will continue this spectroscopic followup campaign in the future, aiming for completeness in the top priority classes.

\section{Preliminary Implications for progenitors}
\label{sec:progenitors}
In this paper, we have presented NIR spectra for 19 RCB stars (eight previously known, 11 newly identified). We have also presented long baseline ($\approx 500$ days) high cadence ($\approx 2$ day) light curves for these stars. In this section, we analyse the NIR spectra to derive diagnostic quantities about the nature of RCB stars. In particular, we focus on photospheric temperatures, oxygen isotope ratios and helium line profiles. We also derive maximum-light pulsation periods for 5 RCB stars. We discuss the implications of these quantities on the progenitors and nature of RCB stars. Although our sample size is not sufficient for any profound insights, the results are indicative of the potential of the NIR campaign. Future observations will increase our sample size and shed more light on the open questions about RCB stars. 

\subsection{Photospheric Temperatures and Isotope ratios}
\subsubsection{SED modeling}
\label{sec:sed_modeling}
We searched archival data from optical, NIR and mid-IR surveys for maximum-light photometric measurements of RCB stars analysed in this paper. For optical wavebands we used data from AAVSO, ZTF and ASAS-SN sky surveys. For NIR wavebands, we compare the 2MASS J-band magnitudes to PGIR lightcurves to identify RCB stars for which the 2MASS measurements correspond to maximum light.  For mid-IR wavebands, we use the WISE magnitudes directly as the mid-IR fluxes of RCB stars do not vary significantly \citep{Feast1997a}. We corrected the WISE 4.6\um magnitudes for saturation effects based on the prescription in \citet{Tisserand2012}. As the distances to these stars are not known accurately ($\approx 30\%$ uncertainty from GAIA-DR2, \citealt{Bailer-Jones2018}), we cannot estimate the line-of-sight extinction in their direction. We restrict our analysis to only those stars that have integrated $E_{B-V}$ values smaller than 1.5 mag. We assume 10\% uncertainties on the photometric measurements to account for variations in the light curve and uncertainties in extinction. We construct maximum light spectral energy distributions (SEDs) for six RCB stars analysed in this paper. This includes five previously known RCB stars $-$ AOHer, NSV11154, ASAS-RCB-21, WISE-J17+, WISE-J19+ and one new RCB WISE-ToI-1213. 

We model the SEDs with the radiative transfer code \texttt{DUSTY} \citep{Ivezic97,Ivezic99,Elitzur01}. In our models, DUSTY is embedded in a Markov Chain Monte Carlo (MCMC) wrapper using the python package \texttt{emcee} \citep{Mackey-Foreman2013}. We modeled the star as a blackbody corrected for molecular absorption. We used synthetic spectra of RCB stars (see Sec. \ref{sec:co_ratios}) to calculate the molecular absorption fractions. The dust is treated as a shell with density profile $\rho \propto r^{-2}$. We fix the outer radius of the shell to be ten times the inner radius \citep{Clayton2011}. We examine optical depth values in the visual band within the range 0 $< \tau_{V} <$ 10 and dust temperatures in the range 300 K $< T_{\mathrm{dust}} < 1200$ K. We assume amorphous carbonaceous dust with a standard MRN size distribution \citep{Mathis1977}. One limitation of this analysis is the assumption of spherical symmetry. Therefore, the estimated dust optical depth could be an over or underestimation depending on the viewing angle. 

\begin{figure*}[t]
    \centering
    \includegraphics[width=\textwidth]{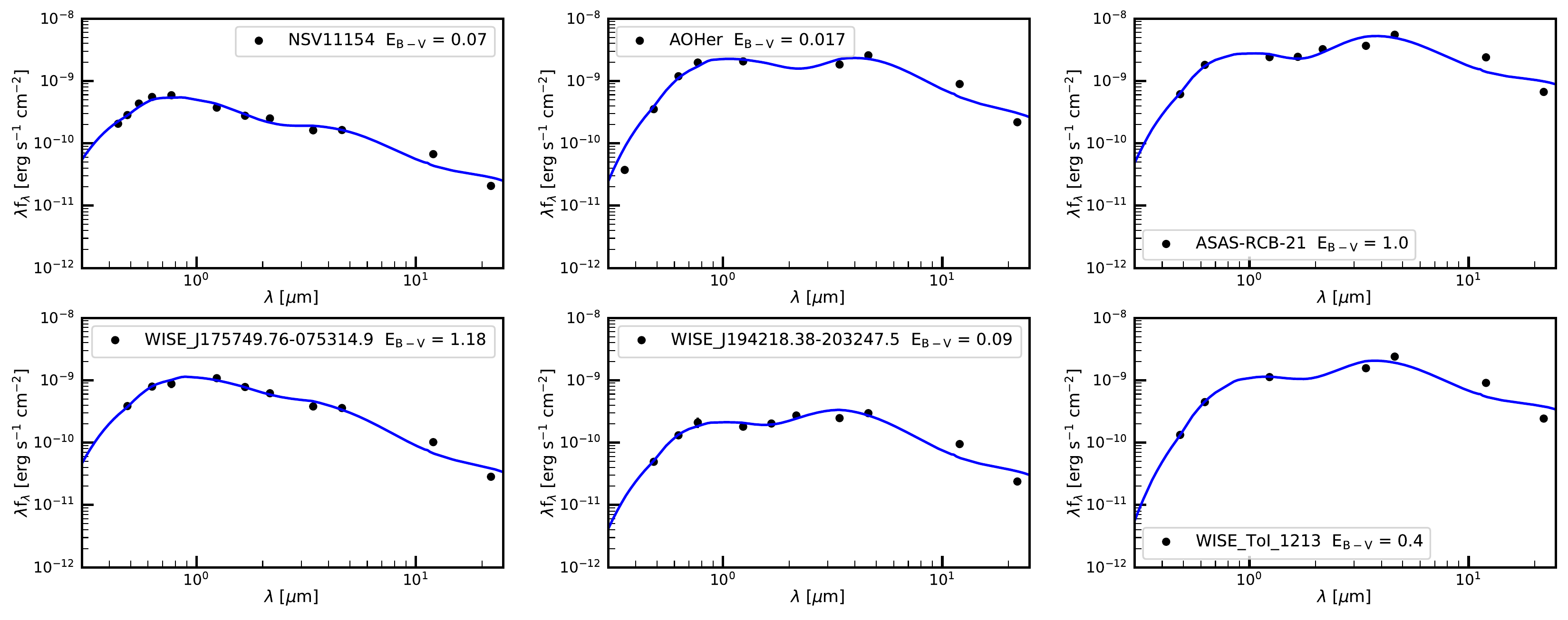}
    \caption{Maximum light spectral energy distributions (SEDs) for six RCB stars from our sample. We plot the best-fit \texttt{DUSTY} models in blue. For each star, we also indicate the assumed values of E$_{B-V}$. The derived model parameters are listed in Table \ref{tab:helium}}
    \label{fig:sed_fits}
\end{figure*}

The SEDs and best fit models from the dust analysis are shown in Figure \ref{fig:sed_fits}. The model parameters and their 1-$\sigma$ uncertainties are listed in Table \ref{tab:helium}. The stars are consistent with temperatures between 4500 and 6000 K, surrounded by dust shells of temperatures between $700-900$ K. With improved distance estimates from the next GAIA data release and additional monitoring from optical and NIR surveys, we will be able to estimate the temperatures for several more RCB stars. Accurate distance measurements will also enable us to derive luminosities of these stars and place them on the Hertzsprung-Russell (HR) diagram.

\subsubsection{NIR spectral modeling}
\label{sec:co_ratios}
The NIR spectra of cold RCB stars provide a host of information about their compositions. Here, we model the CO bandhead in the K band to constrain the temperatures and oxygen isotope ratios of these stars. However, the circumstellar dust shells of RCB stars can contribute significantly to the observed K-band fluxes (up to 80$\%$ of the total flux, \citealt{Tisserand2012}). The observed absorption depth of the CO bands depends largely on the assumed dust contribution. We defer a detailed characterisation of dust shells around these stars to a future paper. Instead, we leave the dust contribution as a free parameter in our model, and report a range of effective temperatures and oxygen isotope ratios for these stars. Secondly, given the medium resolution of our spectra, it is possible that $^{12}$C$^{16}$O bandheads of some RCB stars are saturated \citep{Garcia-Hernandez2009}. It is thus possible that the values of the oxygen isotope ratios are lower limits for some stars. 

We use a grid of Hydrogen deficient spherically symmetric MARCS (Model Atmospheres in Radiative and Convective Scheme) atmospheric models with input compositions characteristic of RCB stars (log $\epsilon$(H) = 7.5, C/He ratio of 1\%, \citealt{Gustafsson1975,Gustafsson08,Bell1976,Plez2008}). We generated the synthetic spectra using the package \texttt{TURBOSPECTRUM} \citep{Alvarez1998}. We set log $g$ = 1.0 and varied the effective temperatures from 4000 to 7500 K in intervals of 250 K. We chose values of 0.5, 1, 2, 5, 10, 20, 50, 500 and infinity for the $^{16}$O/$^{18}$O ratio. We also introduce an additional parameter f$_{\mathrm{dust}} = \frac{F_{\mathrm{shell,K}}}{F_{\mathrm{total,K}}}$ to quantify the dust shell contribution to the K-band flux, and vary it between 0 to 0.8 in steps of 0.1.

\begin{figure*}
    \centering
    \includegraphics[width=\textwidth]{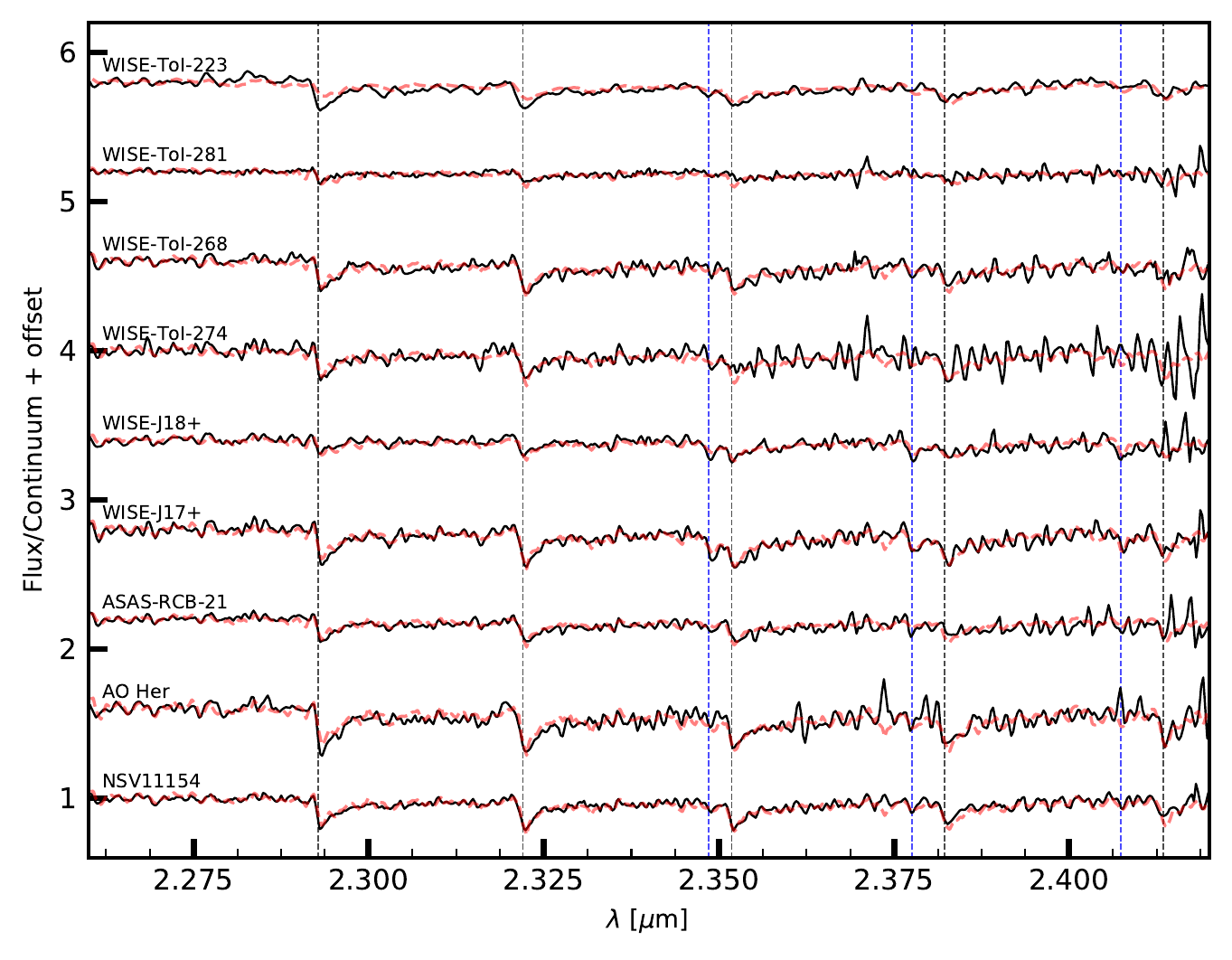}
    \caption{Best-fits to the K band spectra of RCB stars. The continuum normalised spectra are plotted as black solid lines and the best-fit synthetic spectra are plotted as red dashed lines. We mark the positions of $^{12}$C$^{16}$O (black lines) and $^{12}$C$^{18}$O (blue lines) absorption bandheads. The derived ranges of best-fit temperatures and oxygen isotope ratios are listed in Table \ref{tab:helium}.}
    \label{fig:co_bestfits}
\end{figure*}

We fit the synthetic spectra to the continuum normalised NIR spectra in the region 2.26-2.4\um. For each star, we visually selected the spectra that best fit the data for each value of f$_{\mathrm{dust}}$. We used these fits to derive upper limits on the photospheric temperature and a range of O$^{16}$/O$^{18}$ ratios, irrespective of the dust contribution of the star. Figure \ref{fig:co_bestfits} shows the best-fit synthetic spectra together with the continuum normalised NIR spectra. The five RCB stars WISE-J17+, WISE-J18+, WISE-ToI-223, ASAS-RCB-21 and WISE-ToI-268 show the strongest $^{12}$C$^{18}$O absorption features. The first three are consistent with having $^{16}$O/$^{18}$O in the range $0.5-10$ while ASAS-RCB-21 and WISE-ToI-268 are consistent with $^{16}$O/$^{18}$O in $2-50$ and $5-50$ respectively (higher value for higher assumed dust contribution). Using the temperature range for NSV11154 derived in Sec. \ref{sec:sed_modeling}, we constrain the $^{16}$O/$^{18}$O ratio in the range 10-50. For the remaining stars, the $^{12}$C$^{18}$O absorption is weaker. Given the uncertainty in the dust contribution, we cannot put similar constraints on $^{16}$O/$^{18}$O of these stars using our medium resolution spectra. For these stars, we only indicate a lower limit on the oxygen isotope ratios from our fits. Table \ref{tab:helium} lists the derived effective temperature constraints and oxygen isotope ratio ranges. In Section \ref{sec:sed_modeling}, we modeled the SEDs of four of the nine stars analysed here (AOHer, NSV11154, WISE-J17+ and ASAS-RCB-21). We note that the temperatures derived from SED modeling are consistent with constraints derived from NIR spectra. 

The high $^{18}$O abundances in RCB stars could be key to identifying their formation mechanism. The reaction $^{14}$N($\alpha$,$\gamma$) F$^{18}$($\beta^{+} \nu$)$^{18}$O is an important channel for synthesizing $^{18}$O. The cores of intermediate mass stars just before He burning commences can reach the temperatures required for this reaction  \citep{Warner1967}. However, if a star undergoes a final helium flash, the temperature becomes high enough for all $^{18}$O to be destroyed. The FF scenario is thus unlikely to explain the high $^{18}$O abundances \citep{Garcia-Hernandez2009}. On the other hand, dynamically accreting material in mergers of CO-core and He-core white dwarfs can be sites for synthesis of $^{18}$O \citep{Clayton2007}. \citet{Jeffery2011} suggest that even a cold white dwarf merger (i.e., without any nucleosynthesis) could explain large quantities of $^{18}$O in the remnant. These merger models successfully predict the high $^{12}$C/$^{13}$C ratio observed for most RCB stars \citep{Asplund2000}. Most RCB stars are thus thought to be formed from the DD channel. \citet{Garcia-Hernandez2010} suggest that the DD scenario could link RCB stars with Hydrogen-deficient carbon (HdC) stars. This is supported by the high $^{18}$O abundances measured in HdC stars \citep{Clayton2005}. \citet{Garcia-Hernandez2010} note that HdC stars have $^{16}$O/$^{18}$O $< 1$ while RCB stars have $^{16}$O/$^{18}$O $>1$. 

There are a few challenges to the DD picture. Lithium has been detected in the spectra of one HdC \citep{Rao1996} and four RCB stars \citep{Asplund2000}. The RCB star VCrA has a high $^{13}$C abundance \citep{Rao2008}. The presence of Li and $^{13}$C in these stars are inconsistent with DD models, and instead favour an FF origin for them \citep{Asplund2000,Clayton12}. None of these anomalous RCB stars have an $^{18}$O abundance measurement, as most of them are warm.



\begin{figure}
    \centering
    \includegraphics[width=0.5\textwidth]{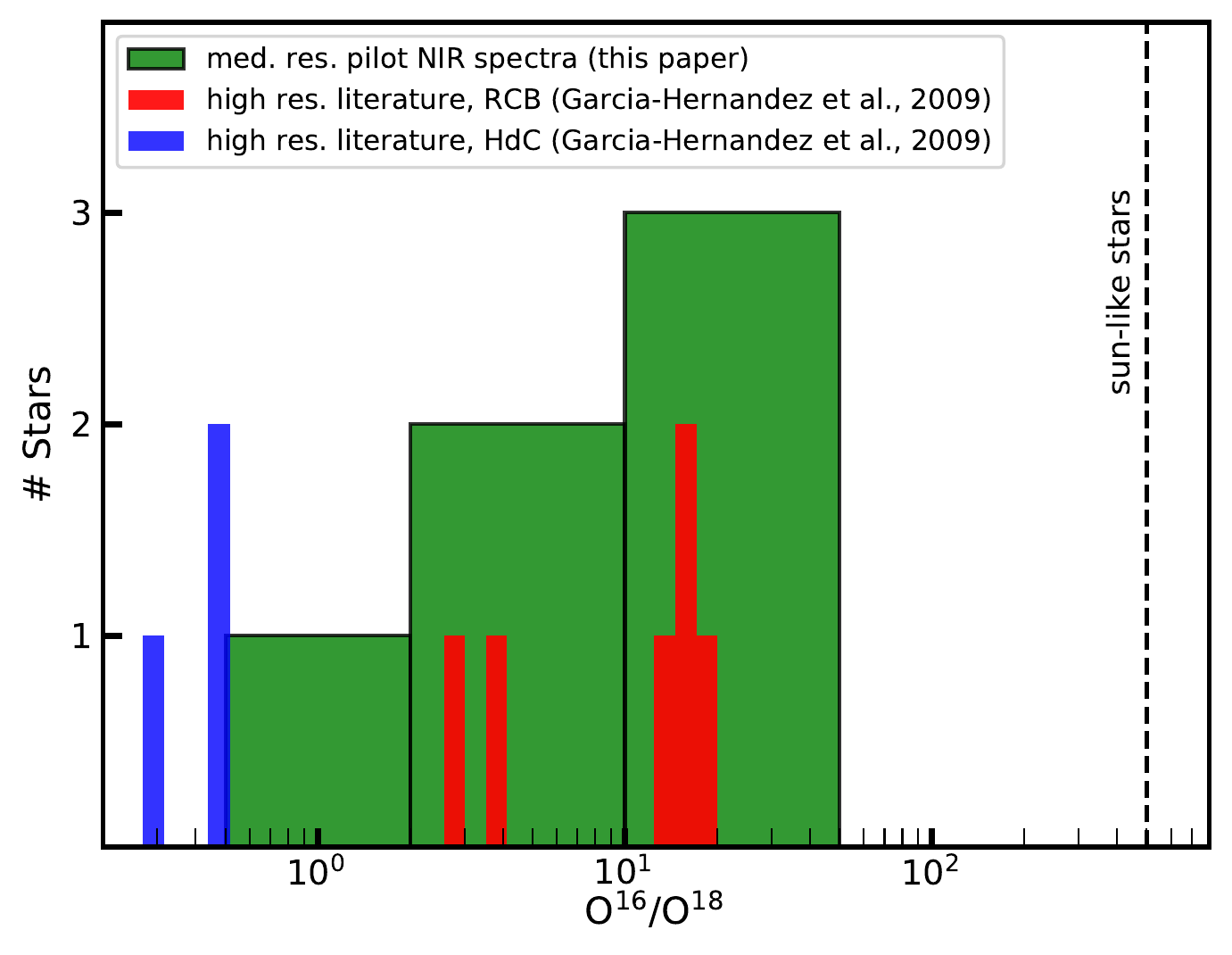}
    \caption{Distribution of $^{16}$O/$^{18}$O ratios of RCB and HdC stars. In red, we plot all 6 previously available measurements of the isotope ratios in RCB stars derived from high resolution spectra \citet{Garcia-Hernandez2009}. In this paper, we report preliminary constraints on $^{16}$O/$^{18}$O for 4 new RCB stars. We indicate these new measurements in green. We note that 5 additional RCB stars reported in this paper show weak $^{12}$C$^{18}$O absorption features, and are consistent with $^{16}$O/$^{18}$O > 10. In blue, we plot the isotope ratios measured for HdC stars, taken from \citet{Garcia-Hernandez2009}. The high $^{18}$O content of RCB stars compared to sun-like stars favours a white dwarf merger origin for them. 
    Further NIR spectroscopic observations that populate this diagram will provide additional clues about the formation channels of RCB stars.}
    \label{fig:o18rats}
\end{figure}

Oxygen isotope ratios have been previously measured for six known Galactic RCB stars and three HdC stars \citep{Clayton2007,Garcia-Hernandez2009,Garcia-Hernandez2010,Bhowmick2018}. We have reported constraints on the ratios of six more RCB stars and provided lower limits for three RCB stars. Figure \ref{fig:o18rats} shows a histogram of $^{16}$O/$^{18}$O ratios for all of these RCB stars. We also indicate ratios for the two HdC stars, and mark the solar $^{16}$O/$^{18}$O value. 
The oxygen isotope ratios in the six RCB stars are at least an order of magnitude lower than the solar value. These stars also do not show any $^{13}$C$^{16}$O features, further indicating that they originate in a WD merger. Additional NIR spectroscopic observations that populate this diagram will provide clues the evolutionary channels of RCB stars. 
In particular, measuring the isotope ratios for anomalous cold RCB stars will help illuminate the multiple possible channels for RCB-formation. 

\subsection{Pulsation Periods and effective temperatures}
At maximum light, some RCB stars are known to pulsate with periods between 40-100 days and amplitudes of a few magnitudes \citep{Lawson1997}. These pulsations can be fairly irregular $-$ the star can exhibit multiple pulsation modes or undergo changes in the dominant period. RCrB has shown pulsations with periods of 33, 44, 52 and 60 days \citep{Lawson1996}. \citet{Saio2008} suggest that these semi-regular or irregular pulsations can be explained by the onset of the strange instability in the remnant of a WD merger. They also find that longer period (P$>100$ days) non-radial modes could be excited in these stars. Using WD merger models, they find that pulsation periods together with effective temperatures of RCB stars can be indicative of their mass. These values have been measured previously for only three RCB stars $-$ RCrB, RY SGr and UW Cen.

The high cadence PGIR light curves are ideal for identifying maximum-light pulsations in RCB stars. Of the 19 RCB stars analysed in this paper, we can measure pulsations for five $-$ AO Her, NSV11154 , V391 Sct, WISE-J18+ and WISE-ToI-8. Four of these are previously known RCB stars but their pulsation periods and temperatures have not been reported. WISE-J18+ has the most well-defined period of 41 days with a Lomb-Scargle (LS) score of 0.4. AO Her shows a period of 56 days (LS score = 0.2), but with significant scatter that could be attributed partly to non-linearilty effects in the PGIR detector. For NSV11154, we detect two distinct pulsation modes. The lightcurve shows one full cycle with a period of 96 days (LS score 0.5) followed by another cycle with a period of 48 days (LS score 0.5) and slightly smaller amplitude. It is possible that the first period is an overtone of the second. V391 Sct shows a full cycle with a period of 124 days (LS score = 0.4). WISE-ToI-8 shows a period of 30 days at low significance (LS score = 0.13).

\begin{figure}
    \centering
    \includegraphics[width=0.5\textwidth]{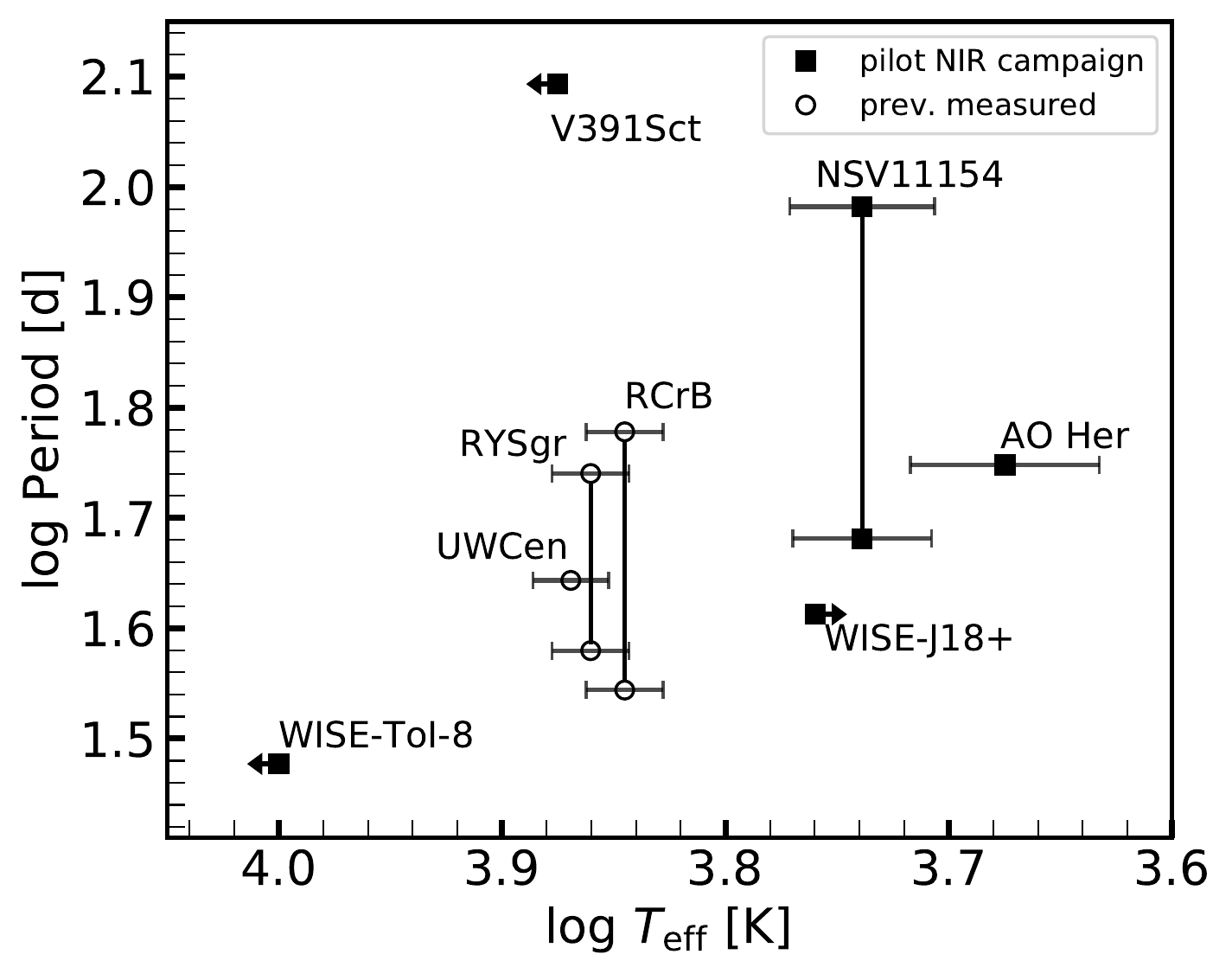}
    \caption{Pulsation periods against effective temperatures for RCB stars. The hollow circles indicate the 3 previously known positions of RCB stars (RCrB, RY Sgr, UW Cen)in this diagram. The solid circles indicate RCB stars analysed in this paper. The positions of RCB stars in this diagram can be compared to white dwarf merger models to infer the masses of the remnants (cf. \citealt{Saio2008}).}
    \label{fig:per-teff}
\end{figure}

Using our NIR spectra, we have put constraints on effective temperatures of 3 of these stars $-$ AO Her, NSV11154 and WISE-J18+ (Table \ref{tab:helium}). WISE-ToI-8 shows several strong emission lines, suggesting that it has $T_{\mathrm{eff}}>10000$K. \citet{Tisserand2013} suggest that V391 Sct has $T_{\mathrm{eff}}>7500$ K based on the presence of metal absorption lines in its spectrum.  
In Figure \ref{fig:per-teff}, we plot the periods of these stars against their effective temperatures. 

\subsection{Dust driven winds in RCB stars}
\label{sec:he_profiles}
\citet{Clayton2003} showed that most (if not all) RCB stars have winds. These winds are thought to be driven by the dust surrounding the star. RCB stars undergo frequent dust formation episodes (a photometric decline results if dust condenses along the line of sight to the star). Radiation pressure from the star can accelerate these dust particles to supersonic velocities. The dust particles drag the gas along with them, giving rise to a wind. Consequently, HdC stars $-$ close relatives of RCB stars that do not undergo major dust formation episodes are not expected to show such winds. 

The helium $\lambda$10830 line is an important tracer of winds in RCB stars for the following reasons $-$ 1) helium is the most abundant elements in RCB stars, 2) The upper level for this line is 20eV above the ground state, and is thus unlikely to be populated by photospheric radiation of F-G type RCB stars. However, this level can be populated by collisional excitation in high velocity winds around these stars. \citet{Clayton2003} first observed this line and noted its ubiquitous presence in RCB stars in their sample. While some stars show a clear p-cygni profile, most show only a strong blueshifted absorption component, as the emission component is affected by other photospheric absorption. They model these profiles to find wind velocities of 200-350 km-s$^{-1}$, with column densities of $10^{12}$cm$^{-2}$. \citet{Clayton2013} observed a larger sample and noted that winds are usually strongest when the RCB is undergoing a decline. 

\begin{figure*}
    \centering
    \includegraphics[width=\textwidth]{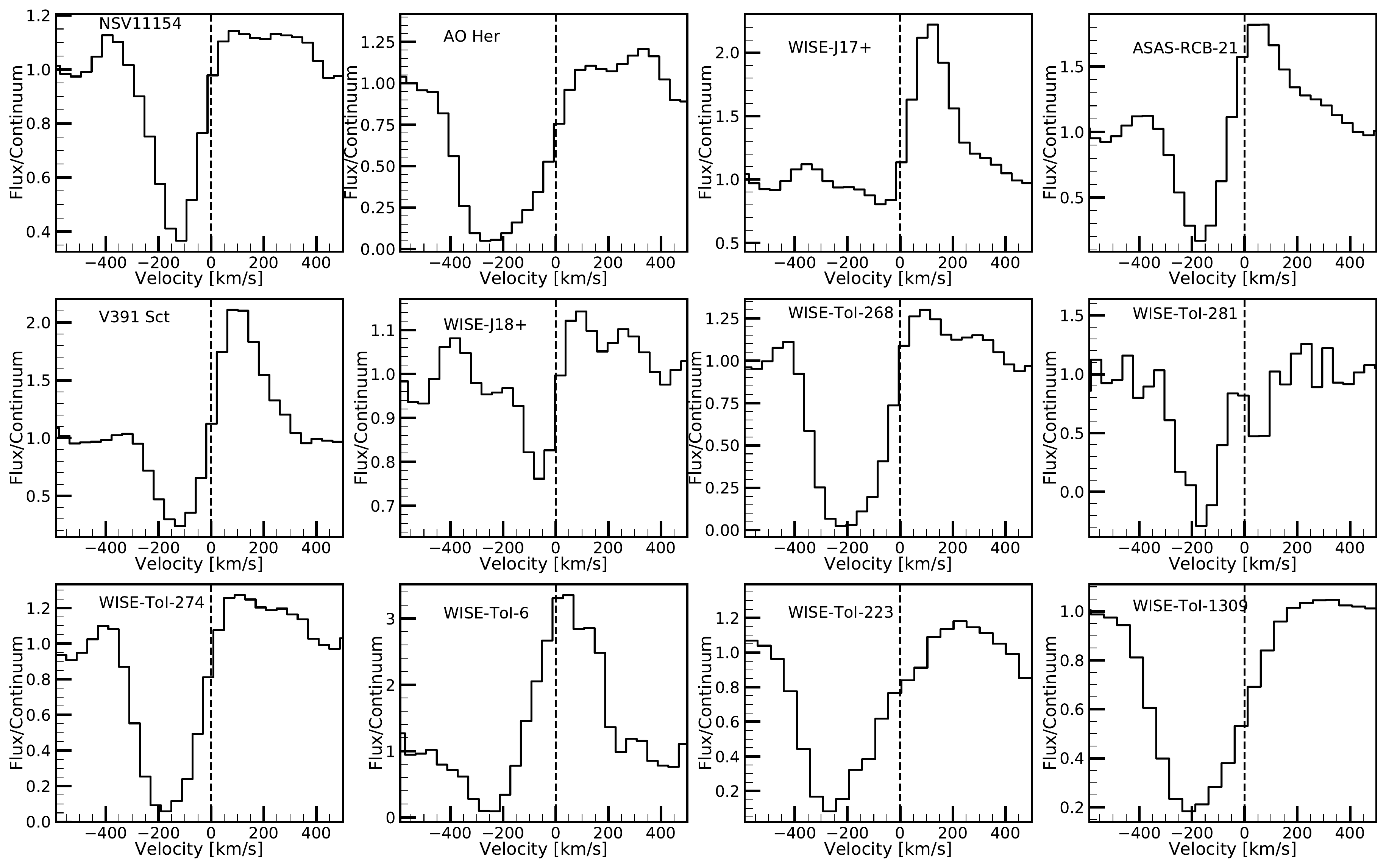}
    \caption{He I $\lambda10830$ line profiles of RCB stars analysed in this paper. The helium line is an almost ubiquitous feature of RCB stars. This line originates in high velocity winds in the atmospheres of these stars. Some RCB stars show a fully developed p-cygni profile, while some stars show only the blueshifted component. In the latter case, other photospheric absorption reduces the emission component. 
    Here, we have excluded the 7 RCB stars that are in a dust enshrouded dip, as we cannot derive radial velocities for them. These stars show the helium line in emission. The absorption component is absent as almost all the continuum is extincted due to dust.}
    \label{fig:He_profiles}
\end{figure*}

Our NIR spectroscopic campaign is ideal to monitor the HeI $\lambda$10833 line in a number of RCB stars in various evolutionary  phases. Similar to \citet{Clayton2003} and \citet{Clayton2013}, we detect strong HeI features in spectra of all RCB  stars analysed in this paper. Of the 19 total RCB stars analysed, 7 are in a deep decline and show the helium line only in emission. As these stars do not have any other spectral features, we cannot measure their radial velocities. We are thus unable to measure wind velocities for these 7 stars. Figure \ref{fig:He_profiles} shows the line profiles for the remaining 12 RCB stars. It is remarkable that some stars show terminal velocities as high as 400-500 km s$^{-1}$. We will present a detailed analysis of the He I profiles in a future paper. 


\begingroup
\renewcommand{\tabcolsep}{3pt}
\begin{table*}
\begin{center}
\begin{minipage}{12cm}
\caption{Properties derived from the spectra of RCB stars. The three stars V391 Sct, WISE-ToI-6 and WISE-ToI-1309 do not show strong CO absorption features.}
\label{tab:helium}
\begin{tabular}{ccccccc}
\hline
\hline
{RCB} & RV $^{a}$ & {$T_{\mathrm{eff}}^{b}$} & {$T_{\mathrm{dust}}^{b}$} & {$\tau_{V}^{b}$} & {$T_{\mathrm{eff}}^{c}$} & O$^{16}$/O$^{18}$ $^{c}$ \\ 
{Name} & {km-s$^{-1}$} & {K} & {K} & & {K} & \\
\hline
AO Her & 500 & 4730$^{+240}_{-200}$ & 830$^{+35}_{-45}$ & 1.25$^{+0.14}_{-0.15}$ & $<5750$ & $>$10  \\
NSV11154 & 300 & 5480$^{+140}_{-170}$ & 770$^{+33}_{-35}$ & 0.38$^{+0.05}_{-0.04}$& 4750$-$5750 & 10$-$50 \\
WISE J17+ & 0 & 4850$^{+160}_{-130}$ & 830$^{+45}_{-60}$ & 0.36$^{+0.08}_{-0.05}$ & $<5750$ & 1$-$5  \\
ASAS-RCB-21 & -50 & 5790$^{+300}_{-390}$ & 800$^{+25}_{-35}$ & 2.10$^{+0.16}_{-0.22}$& $<6250$ & 2$-$50 \\
WISE J19+ & -50 & 5440$^{+350}_{-350}$ & 920$^{+30}_{-50}$ & 1.67$^{+0.16}_{-0.21}$ & &  \\
WISE-ToI-1213 & -50 & 4980$^{+360}_{-400}$ & 790$^{+40}_{-60}$ & 2.21 $^{+0.25}_{-0.46}$ & & \\
WISE J18+ & -150 & & & & $<6250$ & 0.5$-$2  \\
WISE-ToI-223 & -100 & & & & 4750$-$5750 & 2$-$10 \\
WISE-ToI-268 & -150 & & & & $<6000$ & 5$-$50 \\
WISE-ToI-274 & 0 & & & & $<6000$ & $>$10 \\
WISE-ToI-281 & 0 & & & & $<6250$ & $>$5 \\
WISE-ToI-1309 & 0 & .. & .. \\
WISE-ToI-6 & 0 & .. & .. \\
V391 Sct & 0 & .. & .. \\
\hline
\hline
\end{tabular}
\begin{tablenotes} 
\item $a$ : Radial velocity of the RCB star, rounded to nearest 50 km-s$^{-1}$. 
\item $b$ : Quantities derived from modeling of maximum-light SEDs (Sec. \ref{sec:sed_modeling}). 
\item $c$ : Quantities derived from NIR spectral modeling (Sec. \ref{sec:co_ratios}). 

\end{tablenotes}
\end{minipage}
\end{center}
\end{table*}
\endgroup


\subsection{Radial Velocities of RCB stars}
\label{sec:galactic_pop}

\begin{figure}
    \centering
    \includegraphics[width=0.5\textwidth]{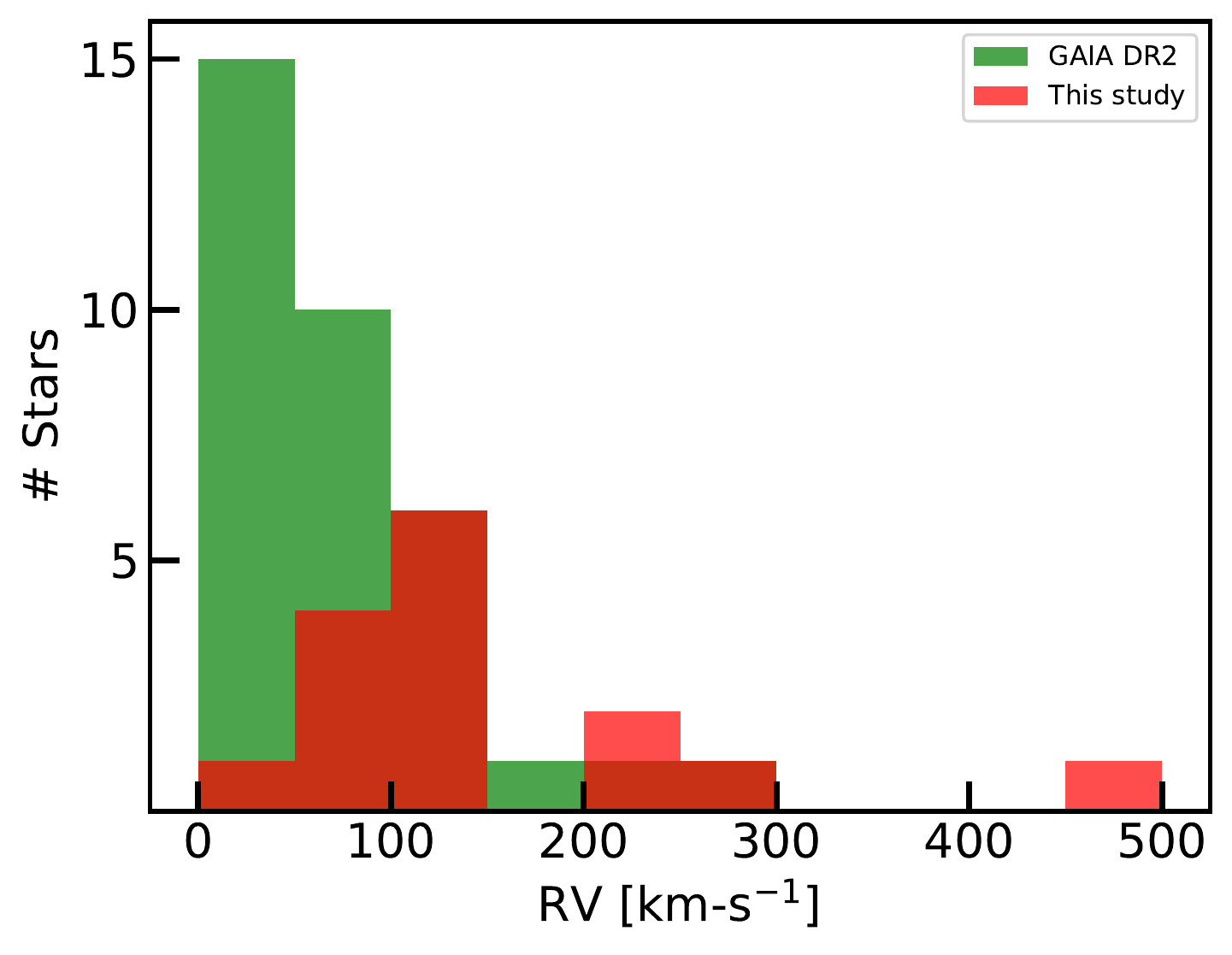}
    \caption{Distribution of the radial velocities of RCB stars. Previously known velocity measurements from GAIA DR2 are plotted in green and measurements from this study are plotted in red. We note that two RCB stars have very high radial velocities ($\gtrapprox 300$ km/s), which could be indicative of a halo population for these objects.}
    \label{fig:hist_rvs}
\end{figure}

The distribution of RCB stars and their velocities will shed further light on their formation channels. Of the 19 RCB stars (11 new + 8 previously known) analysed in this paper, we can measure radial velocities for 11 using C I, CN and CO absorption features in the NIR spectra. Figure \ref{fig:hist_rvs} shows a histogram of the measured velocities. In this figure, we also indicate velocity measurements of other known RCB stars from GAIA DR2. We note that the stars AO Her and NSV11154 have particularly high radial velocities ($\gtrapprox 300$ km-s$^{-1}$). 
Such high velocities could be indicative that these stars could be part of a halo population. Additional measurements that probe the velocity distribution of RCB stars will provide better insights about their formation mechanisms. 

\section{Summary and way forward}
\label{sec:summary}
The population of RCB stars in the Milky Way is largely unexplored $-$ only about a third of Galactic RCB stars are currently known. A systematic census of RCB stars is necessary to shed light on the nature of these mysterious objects. Major advances in this direction were made by \citet{Tisserand2020} with a catalog of 2194 WISE sources expected to contain $\approx 85 \%$ of the RCB stars in the Milky Way. In this paper, we used data from Palomar Gattini IR, a J band time domain survey, to prioritise this catalog for spectroscopic followup. 1209 candidates have PGIR coverage, of which 922 are above the sensitivity of PGIR. Based on their light curves, we assigned a ``lightcurve-based priority" to each of the 922 candidates. The top priority comprises of 149 candidates that show light curves similar to RCB stars. The lowest priority consists of 272 candidates that we classified as long period variables and 23 as RV Tauri stars. Using our lightcurve-based classification, we identified IR color-based criteria that help separate the RCB stars from the LPVs. We then assigned a second ``color-based priority" to each of the 2194 candidates in the catalog. These priorities will aid the process of spectroscopic followup and maximise the yield of RCB stars.

We then obtained NIR spectra for eight known RCB stars and 26 of our top lightcurve-based priority RCB candidates. This was a pilot run for the first NIR spectroscopic campaign to identify RCB stars. We spectroscopically confirmed 11 new RCB stars and analyzed NIR spectral features of these stars. In particular, we report constraints on the effective temperatures and oxygen isotope ratios for these stars. These quantities serve as useful diagnostics about the compositions and formation mechanisms of RCB stars. We derive maximum-light pulsation periods for 5 RCB stars, and plot them in the period-effective temperature space. The position of an RCB star in this phase space can be used to derive the mass of the RCB star. Although the sample analysed in this pilot run is small, it is indicative of the potential of a full NIR spectroscopic campaign for classifying RCB stars. 

\begin{figure}
    \centering
    \includegraphics[width=0.5\textwidth]{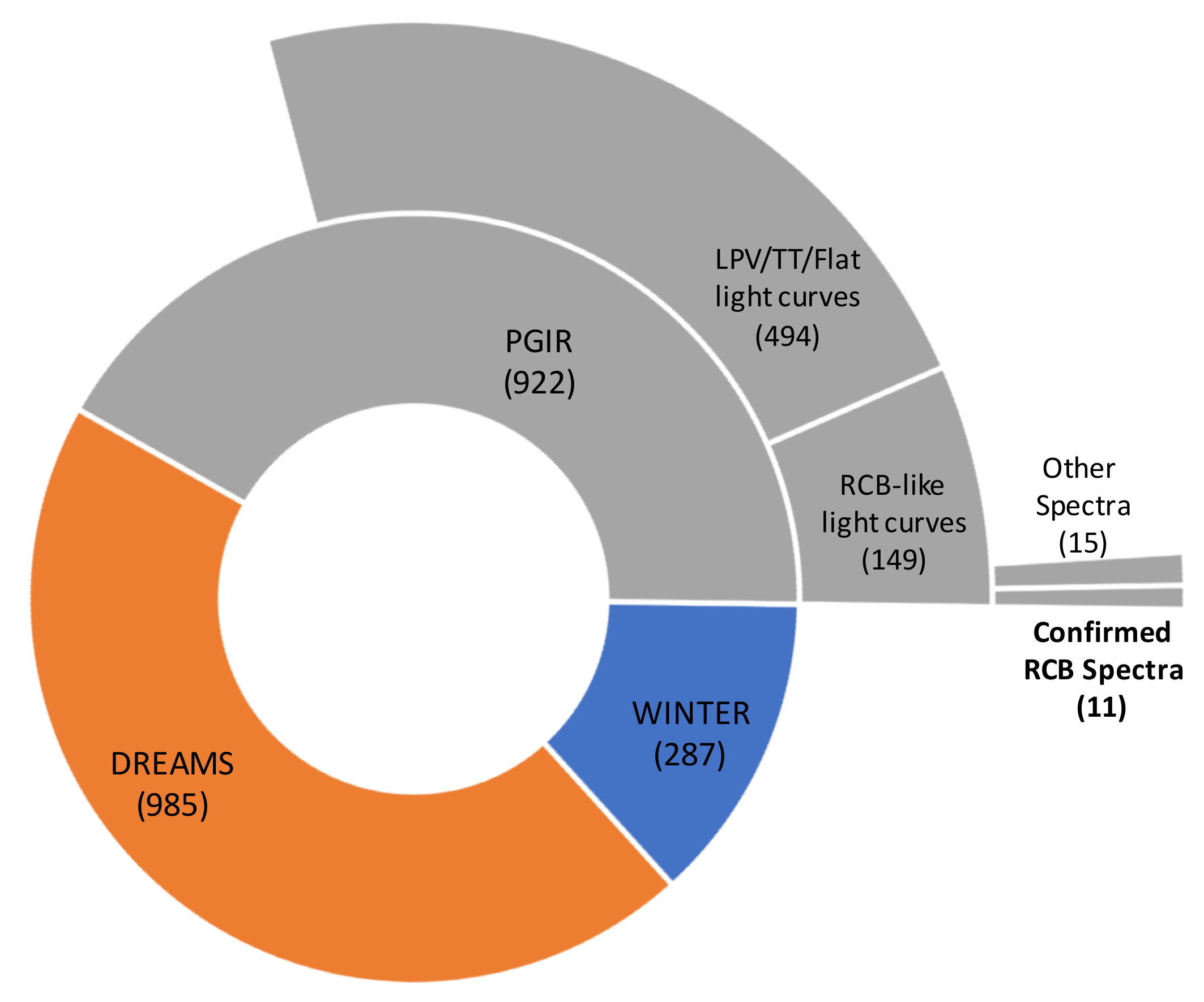}
    \caption{A schematic of our strategy for an NIR survey of Galactic RCB stars focusing on the \citet{Tisserand2020} catalog. 1209 candidates from this catalog are accessible from the northern hemisphere. Of these, 922 candidates are bright enough to be characterised by PGIR. The remaining 287 fainter candidates will be characterised by WINTER $-$ a deeper (J$\approx$20) NIR time domain survey starting December 2021. The candidates in the southern hemisphere will be characterised by DREAMS, an analog of WINTER in Australia.}
    \label{fig:starburst}
\end{figure}

It is evident that NIR searches promise to be productive in identifying and characterising RCB stars. NIR time domain surveys are particularly well suited to discover highly dust enshrouded RCB stars and those lying in highly extincted regions of the Galactic Bulge and plane. The lightcurve and color-based priorities identified in this paper will help identify the remaining RCB stars in the \cite{Tisserand2020} catalog. We aim to complete the spectroscopic followup of top candidates in these priority classes. Further observations from PGIR will help improve the lightcurve-based classifications for the (922) bright candidates visible in the northern hemisphere. The successor to PGIR in the northern hemisphere is Wide-field Infrared Transient Explorer (WINTER, \citealt{Simcoe2019}), a more sensitive J-band time domain survey (lim. J$\approx$ 20 mag). WINTER will start operating at Mt. Palomar by December 2021, and will help characterise the (287) fainter candidates in the northern hemisphere. The Dynamic Red All-sky Monitoring Survey (DREAMS) survey will be an analog of WINTER in the southern hemisphere at the Siding Springs Observatory. DREAMS will help characterise the 985 candidates that are accessible only from the south. Figure \ref{fig:starburst} shows a schematic of this strategy for our NIR census of RCB stars. 

Moreover, systematic blind searches using data from these telescopes will help discover RCB stars outside this catalog. This includes the more exquisite RCB stars with double dust shells around them and DY Per stars$-$ the colder cousins of RCB stars. The NIR census thus promises to uncover the Galactic RCB population.

\bigskip
\section*{Acknowledgements}
Palomar Gattini-IR (PGIR) is generously funded by Caltech, Australian National University, the Mt Cuba Foundation, the Heising Simons Foundation, the Bi- national Science Foundation. PGIR is a collaborative project among Caltech, Australian National University, University of New South Wales, Columbia University and the Weizmann Institute of Science. MMK acknowledges generous support from the David and Lucille Packard Foundation. J. Soon acknowledges the support of an Australian Government Research Training Program (RTP) scholarship. Some of the data presented here were obtained with Visiting Astronomer facility at the Infrared Telescope Facility, which is operated by the University of Hawaii under contract 80HQTR19D0030 with the National Aeronautics and Space Administration.

\bibliography{myreferences}
\bibliographystyle{apj}

\label{lastpage}
\end{document}